\documentclass
[showkeys,showpacs,preprintnumbers,byrevtex]{revtex4} 
\usepackage{amsmath,amsfonts,amssymb,amscd,amsxtra,amsthm}
\usepackage{graphicx}
\usepackage{epstopdf}
\usepackage{bm}
\begin{document}   
\preprint{CYCU-HEP-10-04}
\title{Vector-current correlation and charge separation 
via chiral-magnetic effect}      
%--------------------------------------------------------------------------------------------------
\author{Seung-il Nam}
\email[E-mail: ]{sinam@cycu.edu.tw}
\affiliation{Department of Physics, Chung-Yuan Christian University, Chung-Li 32023, Taiwan} 
%--------------------------------------------------------------------------------------------------
\date{\today}
\begin{abstract}
We investigate the vector-current correlation $\Pi_{\mu\nu}$ (VCC) in the presence of a strong external magnetic field $(\bm{B}=B_{0}\hat{z})$ at low temperature ($T\lesssim T^{\chi}_{c}$) with $C$- and $CP$-violations, indicated by the nonzero chiral-chemical potential ($\mu_{\chi}\ne0$), i.e. the chiral-magnetic effect (CME). For this purpose, we employ the instanton-vacuum configuration at finite $T$ with nonzero topological charge $(Q_{t}\ne0)$. From the numerical calculations, it turns out that the longitudinal component of the connected VCC is liner in $B_{0}$ and shows a bump, representing a corresponding vector meson at $|Q|=(300\sim400)$ MeV for $T=0$. The bump becomes enhanced as $T$ increases and the bump position shifts to a lager $|Q|$ value. In the limit of $|Q|\to0$, the transverse component of the connected VCC disappears, whereas the longitudinal one remains finite and gets insensitive to $B_{0}$  with respect to $T$, due to diluting instanton contributions. Considering a simple collision geometry of HIC and some assumptions on the induced magnetic field and screening effect, we can estimate the charge separation (ChS) as a function of centrality using the present results for VCC. The numerical results show a qualitative agreement with experiments for the Au+Au and Cu+Cu collisions. These results are almost independent on the source of CME, instanton or sphaleron, as long as the CME current is linear in $B_{0}$.
\end{abstract}  
\pacs{12.38.Lg, 14.40.Aq}
\keywords{chiral-magnetic effect, vector-current correlation, charge separation, instanton-vacuum configuration}  
\maketitle
%---------------------------------------------------------------------------------------------------
\section{Introduction}
%---------------------------------------------------------------------------------------------------
QCD at finite $T$ is one of the most important and intriguing subjects for decades. Along with the energetic progresses achieved in heavy-ion collision (HIC) experimental facilities such as the relativistic heavy-ion collision (RHIC) at BNL, theoretical studies to understand QCD at finite $T$ become far more exciting subjects than ever before, especially for the vicinity $\mu\approx0$ and $T\ne0$, resembling the early universe. Beside the fact that the bulk properties of HIC can be interpreted well by relativistic  hydrodynamics~\cite{Hirano:2002ds,Heinz:2001xi}, the microscopic features still seems to be governed by QCD. 

Recently, it was proposed that $P$- and $CP$-violations due to the nontrivial QCD vacuum can be measured experimentally by seeing event-by-event charge separations (ChS) in HIC experiments~\cite{Kharzeev:2004ey,Voloshin:2004vk,Kharzeev:2007jp,Voloshin:2008jx}. Briefly, this phenomenon can be understood as follows: 1) A non-central collision of two heavy nuclei can generate a strong magnetic field, which is perpendicular to the reaction plane. 2) If there is nonzero chiral-chemical potential $\mu_{\chi}\sim N_{R}-N_{L}\ne0$ proportional to nontrivial topological charge $Q_{t}$, indicating tunneling between QCD vacua and $P$- and $CP$-violations as well, 3) quarks with different chiralities (or helicity for massless quarks) move in different directions along the magnetic field. 4) Simultaneously, according to the nonzero $\mu_{\chi}$, there appear electric currents, produced by left-handed or right-handed quarks, resulting in ChS in the measured particles. This is usually called the chiral-magnetic effect (CME). 

Interestingly enough, it was reported that this intresting phenomena, CME was indeed observed by the recent RHIC experiments by STAR collaboration~\cite{:2009txa,:2009uh}. Since ChS is a consequence of event-by-envet fluctuations, if one average it over a space-time volume, it disappears. Thus, a non-vanishing correlation between two particles with electric charges $a$ and $b$, measured in different azimuthal angles ($\Delta\phi$), was suggested~\cite{Voloshin:2004vk,Fukushima:2009ft}:
%EQUATION>>>
\begin{equation}
\label{eq:DEF}
\mathrm{Charge\,separation\,(ChS)}
\equiv\langle\langle \cos(\Delta\phi_{a}+\Delta\phi_{b})\rangle\rangle, 
\end{equation}
%EQUAITON<<<
where $\langle\langle\cdots\rangle\rangle $ denotes an average over the space-time volume. According to this, the experiments observed sizable strengths for ChS for the Au+Au and Cu+Cu collisions in non-central collisions. Theoretical studies also have been done energetically for this direction in various approaches: model-independent analyses on CME~\cite{Fukushima:2008xe,Warringa:2009rw,Asakawa:2010bu}, lattice QCD simulations~\cite{Buividovich:2009wi,Buividovich:2010tn}, effective QCD-like models~\cite{Nam:2009jb,Nam:2009hq,Fukushima:2010fe,Fu:2010rs}, and gauge-gravity duality inspired models~\cite{Yee:2009vw,Gorsky:2010xu,Rebhan:2009vc}, and so on. In what  follows, we briefly introduce these theoretical works. In Refs.~\cite{Fukushima:2008xe}, it was shown that the induced current due to CME along the magnetic field can be derived as:
%EQUATION>>>
\begin{equation}
\label{eq:GJ}
\langle J_{\parallel}\rangle_{\bm{B},\mu_{\chi}} 
=N_{c}N_{f}\frac{\mu_{\chi}B_{0}}{(2\pi^{2})},
\end{equation}
%EQUAITON<<<
where the subscript $\parallel$ stands for that the current is parallel to the magnetic field $\bm{B}=B_{0}\hat{z}$. Here, we set the quark electric charge unity for brevity. As indicated in Ref.~\cite{Fukushima:2008xe} and references therein, this expression is a very general consequence of the axial anomaly in QED. The lattice QCD simulations indicated that the longitudinal component of the current is much larger in comparison to the transverse one. In addition, it was shown that CME becomes insensitive as $T$ increases. These interesting results were also supported and reproduced by the instanton-vacuum configuration model at low $T$~\cite{Nam:2009jb,Nam:2009hq}. On top of the (partial) chiral-symmetry restoration,  deconfinement phase of QCD was also studied with CME employing the Polyakov-loop-augmented Nambu-Jona-Lasinio (PNJL) model~\cite{Fukushima:2010fe,Fu:2010rs}. The chiral-magnetic spiral (CMS) was also investigated with the angular momentum produced in the non-central HIC~\cite{Basar:2010zd}. Some controversial consequences were reported in Refs.~\cite{Yee:2009vw,Gorsky:2010xu,Rebhan:2009vc}, in which the gauge-gravity dual model (AdS/CFT or AdS/QCD based) was exploited. Especially, Ref.~\cite{Yee:2009vw} suggested the negligible CME in the strong-coupling limit. An interpretation were given for this dispute with discussions on back-reactions in Ref.~\cite{Fukushima:2010zz}. Although the experimental observations and theoretical estimations seem promising, it is an urgent task to confirm whether the experimental results are genuine consequences of the nontrivial QCD vacuum effect, i.e. nontrivial correlations between two different vacua $Q_{t}\ne0$ by the possible sources for CME: instanton for low $T$~\cite{Schafer:1996wv,Diakonov:2002fq,Schafer:1995pz} or sphaleron for high $T$~\cite{Arnold:1987zg,Fukugita:1990gb}.  

Considering all of these recent progresses for CME, in the present work, we want to study the vector-current correlation (VCC) in the presence of an external magnetic field and nonzero $Q_{t}$ at low $T$, $T\lesssim T^{\chi}_{c}$, which is the critical $T$ for chiral restoration, employing the instanton-vacuum model as in the previous work~\cite{Nam:2009jb,Nam:2009hq}. This model is characterized by two phenomenological parameters, i.e. average inter-(anti)instanton distance $\bar{R}\approx1$ fm and average instanton size $\bar{\rho}\approx0.3$ fm~\cite{Diakonov:2002fq}. We take into account $T$ modifications on these quantities by introducing the Harrington-Shepard caloron, which is a periodic instanton in the Euclidean temporal direction with trivial holonomy, but without confinement~\cite{Harrington:1976dj,Diakonov:1988my}. As a consequence, we obtain $T$- and momentum-dependent constituent-quark mass, which plays a role of a natural UV regulator. Since the vacuum-expectation value (VEV) of the induced CME current, $\langle J_{\parallel,\perp}\rangle_{\bm{B},\mu_{\chi}}$ was already computed in the previous work~\cite{Nam:2009jb,Nam:2009hq}, we focus to compute the connected VCC in the present work. Using the previous and present results for VCC, we also want to estimate ChS by the following relation~\cite{Kharzeev:2007jp,Fukushima:2009ft}:
%EQUAITON<<<
\begin{equation}
\label{eq:CHA}
\mathrm{ChS}\equiv
\langle\langle \cos(\Delta\phi_{a}+\Delta\phi_{b})\rangle\rangle
\propto\underbrace{\langle J_{\parallel}\rangle^{2}_{\bm{B},\mu_{\chi}}
-\langle J_{\perp}\rangle^{2}_{\bm{B},\mu_{\chi}}}_{\mathrm{Disconnected}}
+\underbrace{
\chi_{\parallel,\bm{B},\mu_{\chi}}
-\chi_{\perp,\bm{B},\mu_{\chi}}}_{\mathrm{Connected}},
\end{equation}
%EQUAITON<<<
where the first and second two terms in the right-hand side come from disconnected and connected quark-loop diagrams of VCC, respectively. Hence, the disconnected VCC can be represented by the squared VEV of the induced current, whereas the connected one by the susceptibility $\chi\sim\partial^{2}\mathcal{S}/\partial A^{2}_{\mu}$. The disconnected diagram contributes to CME primarily, while the connected one corresponds to backgrounds~\cite{Fukushima:2009ft}. It is  worth mentioning that ChS can be sorted to two different kinds: $(a,b)=(\pm,\pm)$ and $(a,b)=(\pm,\mp)$, where $\pm$ stands for the electric charges of the detected particles at different azimuthal angles $\Delta\phi_{a}$ and $\Delta\phi_{b}$ in HIC. Considering the geometrical symmetry of the non-central HIC, ChS for $(+,+)$ and $(-,-)$ will give the same results, whereas that for $(\pm,\mp)$ is different from others in principle. Thus, one can define two different correlations for ChS as same-charge (SCC) and opposite-charge (OCC) correlations, i.e. $\langle\langle \cos(\Delta\phi_{\pm}+\Delta\phi_{\pm})\rangle\rangle$ and $\langle\langle \cos(\Delta\phi_{\pm}+\Delta\phi_{\mp})\rangle\rangle$, respectively.  

From the numerical calculations, it turns out that the longitudinal component of the connected VCC is a liner function in $B_{0}$ and shows a bump, representing a corresponding vector meson at $|Q|=(300\sim400)$ MeV for $T=0$. The bump becomes enhanced as $T$ increases and the bump position shifts to a lager $|Q|$ value. In the limit of $|Q|\to0$, the transverse component of the connected VCC disappears, whereas the longitudinal one remains finite and gets insensitive to $B_{0}$  with respect to $T$, due to diluting instanton contribution. Considering a simple collision geometry of HIC and some assumptions on the induced magnetic field and screening effect of quark-gluon plasma (QGP), we can estimate ChS as a function of the centrality. The numerical results show a qualitatively good agreement with experiments for the Au+Au and Cu+Cu collisions. We observe that the strength of ChS for the Cu+Cu collision is generically larger than that for the Au+Au one as a function of centrality. This tendency is caused by that the probability for the domain with $Q_{t}\ne0$ created in HIC does not depends on centrality, but is proportional to the size of the nucleus inversely~\cite{Kharzeev:2007jp,:2009txa,:2009uh}. Moreover, the absolute value of ChS for OCC becomes smaller than that for SCC, due to the screening effect. We note that these results are almost independent from what the source of CME is, instanton or sphaleron, as long as the CME current being linear in $\mu_{\chi}B_{0}$. 

We organize the present work as follows: In Section II, we briefly introduce the instanton-vacuum model and define a quantity $\delta$, which indicates the strength of $P$- and $CP$-violations with nonzero $Q_{t}$. We obtain an effective quark-instanton action as a function of $\delta$. We define connected VCC and corresponding scalar VCC in the presence of the external magnetic field and $\delta$ in Section III. In Section IV, $T$ modifications on the instanton parameters $\bar{R}$ and $\bar{\rho}$ are taken into account in terms of the Harrington-Shepard caloron, and the constituent-quark mass is then defined as a function of momentum and $T$. Considering all the ingredients discussed in the previous Sections, we write expressions for VCC relating to CME as functions of $T$, $\delta$, and $B_{0}$ in Section V. In Section VI, we discuss a  simple collision geometry of HIC. Numerical results for VCC and ChS are given in Section VII with discussions. The final Section is devoted for summary and conclusion of the present work.         
%---------------------------------------------------------------------------------------------------
\section{Effective action from the instanton vacuum with $P$ and $CP$ violations}
%---------------------------------------------------------------------------------------------------
In this Section, we briefly introduce a $P$- and $CP$-violating effective action $\mathcal{S}_{\mathrm{eff}}$, derived by Diakonov et al. from the instanton-vacuum configuration in the large $N_{c}$ limit at zero $T$~\cite{Diakonov:1995qy}. Employing a dilute grand canonical ensemble of the (anti)instantons with finite instanton-number fluctuations, $\Delta\equiv N_{+}-N_{-}\ne0$, which corresponds to a $CP$-violating vacuum, but a fixed total number of the pseudo-particles $N_{+}+N_{-}=N$, $\mathcal{S}_{\mathrm{eff}}$ can be written in momentum space with the Euclidean metric as follows:
%EQUATION>>>
\begin{eqnarray}
\label{eq:EA}
\mathcal{S}_{\mathrm{eff}}
&=&\mathcal{C}+\frac{N_{+}}{V}\ln\lambda_{+}
+\frac{N_{-}}{V}\ln\lambda_{-}
-\frac{mN_{c}}{4\pi^{2}\bar{\rho}^{2}}(\lambda_{+}+\lambda_{-})
\cr
&-&N_{c}\int\frac{d^{4}k}{(2\pi)^{4}}\mathrm{Tr}_{\gamma}\ln
\left[\frac{\rlap{/}{k}-\frac{i}{2}
[\lambda_{+}(1+\gamma_{5})+\lambda_{-}(1-\gamma_{5})]F^{2}(k)}
{\rlap{/}{k}-im}\right],
\end{eqnarray}
%EQUAITON<<<
where we have used $N_{f}=1$ for simplicity. However, the extension to an arbitrary $N_{f}$ is just straightforward. $\mathcal{C}$ stands for an irrelevant constant for further investigations, whereas $N_{\pm}/V$ for the (anti)instanton packing fraction proportional to the inverse of the average inter-(anti)instanton distance $1/\bar{R}^{4}\approx(200\,\mathrm{MeV})^{4}$. $\lambda_{\pm}$ denotes a Lagrangian multiplier, which was employed to exponentiate the $2N_{f}$-'t Hooft interaction in the effective action~\cite{Diakonov:2002fq}. The average instanton size in the dilute instanton ensemble is assigned as $1/\bar{\rho}\approx600$ MeV, while $m$ indicates a small but finite current-quark mass for the SU(2) light-flavor sector ($m\to0$). $F(k)$ denotes the quark form factor originating from the non-local quark-instanton interactions and is defined as
%EQUATION>>>
\begin{equation}
\label{eq:FF}
F(k)=\frac{t}{2}{\rho}\left[I_{0}(t)K_{1}(t)-I_{1}(t)K_{0}(t)
-\frac{1}{t}I_{1}(t)K_{1}(t) \right],\,\,\,\,t=\frac{|k|\bar{\rho}}{2},
\end{equation}
%EQUAITON<<<
where $I_{n}$ and $K_{n}$ stand for the modified Bessel functions. We, however, will employ a parameterization of this from factor for convenience in the numerical calculations in what follows. 

From the effective action, we can obtain the following two self-consistent (saddle-point) equations with respect to $\lambda_{\pm}$:
%EQUATION>>>
\begin{eqnarray}
\label{eq:SDP}
\lambda_{\pm}
\frac{\partial \mathcal{S}_{\mathrm{eff}}}{\partial \lambda_{\pm}}
&=&\frac{N_{\pm}}{V}-\frac{\lambda_{\pm}mN_{c}}{4\pi^{2}\bar{\rho}^{2}}
+N_{c}\int\frac{d^{4}k}{(2\pi)^{4}}\mathrm{Tr}_{\gamma}
\left[\frac{\frac{i\lambda_{\pm}}{2}
(1\pm\gamma_{5})F^{2}(k)}{\rlap{/}{k}-\frac{i}{2}
[\lambda_{+}(1+\gamma_{5})+\lambda_{-}(1-\gamma_{5})]F^{2}(k)}\right]
\cr
&=&\frac{N_{\pm}}{V}-\frac{(1\pm\delta)M_{0}mN_{c}}
{4\pi^{2}\bar{\rho}^{2}}
-N_{c}\int\frac{d^{4}k}{(2\pi)^{4}}\mathrm{Tr}_{\gamma}
\left[\frac{\frac{1}{2}(1\pm\gamma_{5})
(1+\delta\gamma_{5})^{2}M^2}
{k^{2}+(1+\delta\gamma_{5})^{2}M^2}\right]=0,
\end{eqnarray}
%EQUAITON<<<
where $\lambda_{\pm}$ is approximated as $M_{0}(1\pm\delta)$ in the last line of Eq.~(\ref{eq:SDP}) accounting for $\Delta\ll N$ in the thermodynamic limit~\cite{Diakonov:1995qy}. The momentum-dependent constituent-quark mass is defined as $M_{K}=M_{0}F^{2}(k)$~\cite{Diakonov:2002fq}. By adding and subtracting the instanton $(+)$ and anti-instanton ($-$) contributions in Eq.~(\ref{eq:SDP}), we arrive at
%EQUATION>>>
\begin{equation}
\label{eq:NOV}
\frac{N}{V}-\frac{mM_{0}N_{c}}
{2\pi^{2}\bar{\rho}^{2}}\approx4N_{c}\int\frac{d^{4}k}{(2\pi)^{4}}
\frac{(1+\delta^{2})M^{2}}
{k^{2}+(1+\delta^{2})M^{2}},
\end{equation}
%EQUAITON<<<
%EQUATION>>>
\begin{equation}
\label{eq:DOV}
\frac{\Delta}{V}-\frac{\delta mM_{0}N_{c}}
{2\pi^{2}\bar{\rho}^{2}}\approx8N_{c}\int\frac{d^{4}k}{(2\pi)^{4}}
\frac{\delta M^{2}}
{k^{2}+(1+\delta^{2})M^{2}}.
\end{equation}
%EQUAITON<<<
Taking into account $\delta\ll1$, $\Delta\ll N$, and using Eq.~(\ref{eq:DOV}), we can obtain an expression for $\delta$ as a function of relevant parameters:
%EQUATION>>>
\begin{equation}
\label{eq:DEL}
\delta=\left(\frac{2\pi^{2}\bar{\rho}^{2}}{mM_{0}N_{c}}\right)
\frac{\Delta}{V}.
\end{equation}
%EQUAITON<<<
This equation tells us that $\delta$ contains the information on the instanton-number fluctuation $\Delta$ at a certain scale $\bar{\rho}$, which is about $600$ MeV in the present framework for vacuum. Taking into account all the ingredients discussed so far, finally, we can write the relevant effective action with $\Delta\ne0$ for further investigations:
%EQUATION>>>,
\begin{eqnarray}
\label{eq:EA2}
\mathcal{S}_{\mathrm{eff}}
&=&-\int\frac{d^4k}{(2\pi)^4}\mathrm{Tr}_{c,f,\gamma}\ln
\left[\frac{\rlap{/}{k}-i(1+\delta\gamma_{5})M_{K}}
{\rlap{/}{k}-im}\right],
\end{eqnarray}
%EQUAITON<<<
where the $\mathrm{Tr}_{c,f,\gamma}$ denotes the trace over color, flavor  and Lorentz indices. 

Now, we are in a position to discuss the relation between the nontrivial topological charge $Q_{\mathrm{t}}$, as a source of CME, and the instanton number fluctuation $\Delta$. According to the axial Ward-Takahashi identity~\cite{Kharzeev:2007jp}, $Q_{\mathrm{t}}$ is proportional to the number difference between the chirally left- and right-handed quarks, $Q_{\mathrm{t}}\propto N_{R}-N_{L}$. Hence, nonzero $Q_{\mathrm{t}}$ indicates the chirality flip. Note that, similarly, if a chirally left-handed quark is scattered from an instanton to an anti-instanton, the quark helicity is flipped to the right-handed one, and vice versa. This means that the nonzero $\Delta$ results in $N_{R}-N_{L}\ne0$. In this way, $Q_{\mathrm{t}}$ can be considered to be proportional to $\Delta$: $Q_{\mathrm{t}}\sim\Delta$~\cite{Schafer:1996wv,Diakonov:2002fq}. As a consequence, we can study CME using the effective action in Eq.~(\ref{eq:EA2}) as a function of $Q_{\mathrm{t}}$, more explicitly $\delta\propto\Delta$. Since $\delta$ stands for the symmetry breaking, its order of strength must be very small, corresponding to the order of $CP$ violation. In Table~\ref{TABLE0}, we summarize the numerical values for the relevant inputs. 
%EQUATION>>>
\begin{table}[b]
\begin{tabular}{c|c|c|c|c}
$\bar{R}$
&$\bar{\rho}$
&$(N/V)^{1/4}$
&$M_{0}$
&$m$\\
\hline
$1$ fm&$1/3$ fm&$197$ MeV
&$325$ MeV&$5$ MeV\\
\end{tabular}
\caption{Instanton parameters, constituent and current quark masses in vacuum.}
\label{TABLE0}
\end{table}
%EQUAITON<<<
%---------------------------------------------------------------------------------------------------
\section{Vector-current correlation in the presence of $B_{0}$ and $\delta$}
%--------------------------------------------------------------------------------------------------
%FIGURE>>>
\begin{figure}[t]
\includegraphics[width=7.5cm]{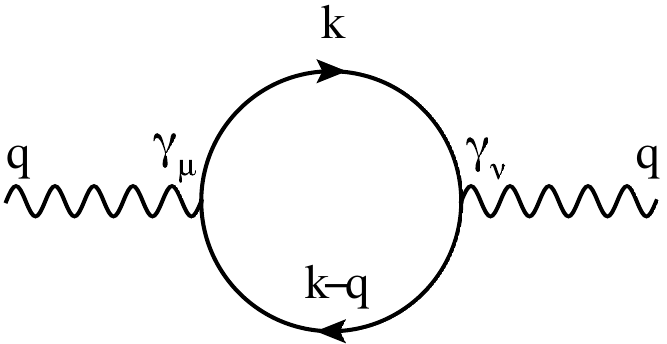}
\caption{Connected vector-current correlation (VCC) in the leading contribution. The wavy and solid lines indicate the vector-particle and quark lines, respectively.}       
\label{FIG0}
\end{figure}
%FIGURE<<<

In this Section, we discuss and derive VCC from the effective action in Eq.~(\ref{eq:EA2}).  First, we write a local vector-current operator in Euclidean space as follows: 
%EQUATION>>>
\begin{equation}
\label{eq:VC}
\mathcal{V}_{\mu}^{a}(x)
=-iq^{\dagger}(x)\gamma_{\mu}\frac{\tau^{a}}{2}q(x),
\end{equation}
%EQUAITON<<<
where $q(x)$ and $\tau^{a}$ stand for a quark field and SU(2) isospin matrix, respectively. Using this definition, we can write VCC ($\mathcal{V}^{a}_{\mu}$), represented by VEV of the time ordering of the two vector-current operators:
%EQUATION>>>
\begin{equation}
\label{eq:PIT}
\Pi^{ab}_{\mu\nu}(q)=-\int d^{4}x\,e^{iq\cdot x}\langle0|\mathcal{T}\left[\mathcal{V}^{a}_{\mu}(x), \mathcal{V}^{b}_{\nu}(0)\right]|0 \rangle, 
\end{equation}
%EQUAITON<<<
where superscript $a$ and momentum $q$ denote an isospin index and the momentum of a corresponding vector meson, respectively. Taking into account vector-current conservation, one obtains an expression for a scalar VCC from Eq.~(\ref{eq:PIT}) as follows:
%EQUATION>>>
\begin{equation}
\label{eq:PIS}
\Pi^{ab}_{\mu\nu}(q)
=\left(\delta_{\mu\nu}-\frac{q^{\mu}q^{\nu}}{q^{2}} \right)\Pi^{ab}(q).
\end{equation}
%EQUAITON<<<
According to this, one is lead to a simple relation between the tensor and scalar VCC:
%EQUATION>>>
\begin{equation}
\label{eq:PITPIS}
\delta^{\mu\nu}\Pi^{ab}_{\mu\nu}(q)=3\Pi^{ab}(q).
\end{equation}
%EQUAITON<<<
Substituting Eq.~(\ref{eq:PITPIS}) into Eq.~(\ref{eq:PIT}), we have the following equation:
%EQUATION>>>
\begin{equation}
\label{eq:PIS2}
\Pi^{ab}(q)=-\frac{\delta^{\mu\nu}}{3}\int d^{4}x\,e^{iq\cdot x}
\langle0|\mathcal{T}\left[\mathcal{V}^{a}_{\mu}(x), 
\mathcal{V}^{b}_{\nu}(0)\right]|0 \rangle, 
\end{equation}
%EQUAITON<<<

In order to compute VCC in Eqs.~(\ref{eq:PIS}) and (\ref{eq:PIS2}), we rewrite the effective action with an external-source vector field $\mathcal{V}^{a}_{\mu}$, using Eq.~(\ref{eq:EA2}):
%EQUATION>>>
\begin{eqnarray}
\label{eq:EA3}
\mathcal{S}_{\mathrm{eff}}
&=&-\int\frac{d^4k}{(2\pi)^4}\mathrm{Tr}_{c,f,\gamma}\ln
\left[\rlap{\,/}{K}-i(1+\delta\gamma_{5})M_{K}
+\rlap{/}{\mathcal{V}}^{\alpha}\tau_{\alpha}\right],
\end{eqnarray}
%EQUAITON<<<
where we have set the current-quark mass zero in the effective action, since it is irrelevant in further discussions. $K_{\mu}$ indicates covariant quark momentum, gauged by a photon field as $k_{\mu}+A_{\mu}$, in which we set the quark electric charge unity for convenience. The electromagnetic (EM) field configuration is chosen to make the external magnetic field along the $z$ direction, $\bm{B}=B_{0}\hat{z}$, which is set to be perpendicular to the collision plane of HIC: 
%EQUATION>>>
\begin{equation}
\label{eq:A}
A_{\mu}=A^{\mathrm{clas}}_{\mu}+A^{\mathrm{fluc}}_{\mu}
=\left(-\frac{B_{0}}{2}x_{2},\frac{B_{0}}{2}x_{1},0,0 \right)
+\left(\bm{A},A_{4} \right).
\end{equation}
%EQUAITON<<<
Here the fluctuations are assumed to be static and far smaller than the model scale: $(|\bm{A}|,A_{4})\ll\Lambda\approx1/\bar{\rho}$. According to this configuration, only $F_{12}$ and its dual $\tilde{F}_{34}$ will survive in the EM field-strength tensor. Then VCC can be easily evaluated by performing functional derivatives of Eq.~(\ref{eq:EA3}) with respect to $\mathcal{V}^{a}_{\mu}(x)$ and $\mathcal{V}^{b}_{\mu}(0)$, resulting in
%EQUATION>>>
\begin{eqnarray}
\label{eq:JKI}
\langle0|\mathcal{T}\left[\mathcal{V}^{a}_{\mu}(x), 
\mathcal{V}^{b}_{\nu}(0)\right]|0\rangle=
i\int\frac{d^4p}{(2\pi)^4}e^{-ip\cdot x}\mathcal{F}^{ab}_{\mu\nu}(p).
\end{eqnarray}
%EQUAITON<<<

In order to induce an external electromagnetic field to the quark-instanton system as Eq.~(\ref{eq:A}), we have used the linear Schwinger method~\cite{Schwinger:1951nm}. $F^{ab}_{\mu\nu}(p)$ for {\it connected} VCC is then given by
%EQUATION>>>
\begin{equation}
\label{eq:FAB}
\mathcal{F}^{ab}_{\mu\nu}(p)=\frac{i\delta^{ab}}{2}\int\frac{d^4k}{(2\pi)^4}
\mathrm{Tr}_{c,f,\gamma}
\left[\frac{1}{\rlap{\,/}{K}-i(1+\delta\gamma_{5})M_{k}} \gamma_{\nu}\frac{1}{\rlap{\,/}{K}-\rlap{/}{p}-i(1+\delta\gamma_{5})M_{k-p}}\gamma_{\mu}\right].
\end{equation}
%EQUAITON<<<
Here, we have used $\mathrm{Tr}[\tau^{a}\tau^{b}]=2\delta^{ab}$. The gauged quark propagator inside the square bracket of the right-hand side can be expanded in terms of $\delta$ and $A_{\mu}$, and we obtain
%EQUATION>>>
\begin{eqnarray}
\label{eq:DEEX}
\frac{1}{\rlap{/}{K}-i(1+\delta\gamma_{5})M(K)}
&\approx&
\frac{\rlap{/}{k}+\rlap{/}{A}
+i(1+\delta\gamma_{5})\left[M+\frac{1}{2}\bar{M}
(\sigma\cdot F)\right]}
{k^{2}+(1+\delta^{2})M^{2}}
\cr
&\times&
\left[1-\frac{\tilde{M}(\sigma\cdot F)+i\hat{M}(k)\gamma_{\mu}K_{\nu}F_{\mu\nu}-2i\delta M\gamma_{5}\rlap{/}{K}}{k^{2}+(1+\delta^{2})M^{2}}\right],
\end{eqnarray}
%EQUAITON<<<
where $\sigma\cdot F=\sigma_{\mu\nu}F^{\mu\nu}$ with the antisymmetric tensor $\sigma_{\mu\nu}=i(\gamma_{\mu}\gamma_{nu}-\gamma_{\nu}\gamma_{\mu})/2$ and the EM field strength tensor $F^{\mu\nu}=\partial_{\mu} A_{\nu}-\partial_{\nu} A_{\mu}$. The relevant mass functions in Eq.~(\ref{eq:DEEX}) are also defined as:
%EQUATION>>>
\begin{equation}
\label{eq:MF}
M_{k}=M_{0}\left(\frac{2}{2+k^{2}\bar{\rho}^{2}} \right)^{2},\,\,\,\,
\bar{M}_{k}=-\frac{8M_{0}\bar{\rho}^{2}}{(2+k^{2}\bar{\rho}^{2})^{3}},\,\,\,\,
\tilde{M}_{k}=\frac{1}{2}+M_{k}\bar{M}_{k},\,\,\,\,\hat{M}_{k}=4i\bar{M}_{k}.
\end{equation}
%EQUAITON<<<
Using Eq.~(\ref{eq:DEEX}) and collecting the terms of $\mathcal{O}(\delta)$, the leading contributions for the transverse $(\mu,\nu=1,2)$, longitudinal $(\mu,\nu=3)$, and temporal $(\mu,\nu=4)$ components of Eq.~(\ref{eq:FAB}) can be obtained as follows:
%EQUATION>>>
\begin{eqnarray}
\label{eq:F1234}
\mathcal{F}^{ab}_{11,22}(p)&=&\int\frac{d^4k}{(2\pi)^4}
\frac{4i\delta^{ab}\delta B_{0}N_{c}N_{f}}{(k^{2}+M^{2}_{k})
[(k-p)^{2}+M^{2}_{k-p}]}
\left[\frac{M_{k}\bar{M}_{k}}{k^{2}+M^{2}_{k}}
-\frac{M_{k-p}\bar{M}_{k-p}}{(k-p)^{2}+M^{2}_{k-p}} 
\right]\epsilon^{12\sigma\rho}k_{\sigma}(k-p)_{\rho},
\cr
\mathcal{F}^{ab}_{33}(p)&=&\int\frac{d^4k}{(2\pi)^4}
\frac{4i\delta^{ab}\delta B_{0}N_{c}N_{f}}{(k^{2}+M^{2}_{k})
[(k-p)^{2}+M^{2}_{k-p}]}
\left[\frac{M_{k}\bar{M}_{k}}{k^{2}+M^{2}_{k}}
+\frac{M_{k-p}\bar{M}_{k-p}}{(k-p)^{2}+M^{2}_{k-p}} 
\right][\epsilon^{12\sigma\rho}k_{\sigma}(k-p)_{\rho}+k_{4}p_{3})]
\cr
\mathcal{F}^{ab}_{44}(p)&=&\int\frac{d^4k}{(2\pi)^4}
\frac{4i\delta^{ab}\delta B_{0}N_{c}N_{f}}{(k^{2}+M^{2}_{k})
[(k-p)^{2}+M^{2}_{k-p}]}
\left[\frac{M_{k}\bar{M}_{k}}{k^{2}+M^{2}_{k}}
+\frac{M_{k-p}\bar{M}_{k-p}}{(k-p)^{2}+M^{2}_{k-p}} 
\right][\epsilon^{12\sigma\rho}k_{\sigma}(k-p)_{\rho}-k_{3}p_{4})],
\end{eqnarray}
%EQUAITON<<<
where we have used the fact that, due to our choice of the EM-field configuration in Eq.~(\ref{eq:A}), only $F_{12}=B_{0}$ survives among the field-strength tensor. It is clear that the sign difference in the square bracket in the right-hand side makes the longitudinal and temporal contributions $\mathcal{F}^{ab}_{33,44}$ are significantly larger than that for the transverse ones, which is a general consequence of CME as indicated in Refs.~\cite{Nam:2009jb}. The term with $\epsilon$ tensor can be evaluated further as
%EQUATION>>>
\begin{equation}
\label{eq:AASS}
\epsilon^{12\sigma\rho}k_{\sigma}(k-p)_{\rho}=k_{3}(k_{4}-p_{4})-k_{4}(k_{3}-p_{3})=k_{4}p_{3}-k_{3}p_{4}.
\end{equation}
%EQUAITON<<<
Finally, we arrive at the following expressions for connected VCC and corresponding scalar VCC using Eq.~(\ref{eq:PIS}) as follows:
%EQUATION>>>
\begin{eqnarray}
\label{eq:SPEC}
\Pi^{ab}_{\mu\nu}(q)&=&-i
\int d^{4}x\int\frac{d^4p}{(2\pi)^4}e^{-i(p-q)\cdot x}
\mathcal{F}^{ab}_{\mu\nu}(p)=-i\mathcal{F}^{ab}_{\mu\nu}(q),
\cr
\Pi^{ab}(q)&=&-\frac{i\delta^{\mu\nu}}{3}
\int d^{4}x\int\frac{d^4p}{(2\pi)^4}e^{-i(p-q)\cdot x}
\mathcal{F}^{ab}_{\mu\nu}(p)=-\frac{i}{3}\sum_{n=1}^{4}\mathcal{F}^{ab}_{nn}(q)=-\frac{i}{3}
\left[2\mathcal{F}^{ab}_{11}(q)+\mathcal{F}^{ab}_{33}(q)+\mathcal{F}^{ab}_{44}(q) \right].
\end{eqnarray}
%EQUAITON<<<
Since $p$ and $k$ are the integral variables as in Eqs.~(\ref{eq:F1234}) and (\ref{eq:SPEC}), the interchange $k\leftrightarrow p$ does not make any difference in the absolute values of the longitudinal and temporal components, i.e. $|\Pi^{ab}_{33}(q)|=|\Pi^{ab}_{44}(q)|$. However, as will be shown in Section V, they become differing from each other at finite $T$, due to the breakdown of Lorentz invariance, since the fourth component of four momenta becomes periodic in the temporal direction. 

%---------------------------------------------------------------------------------------------------
\section{Instanton effects at finite temperature}
%---------------------------------------------------------------------------------------------------
To investigate the physical quantities in hand at low but finite $T$ ($T\lesssim T^{\chi}_{c}$), we want to discuss briefly how to modify the instanton variables, $\bar{\rho}$ and $\bar{R}$ at finite $T$. We will follow our previous work~\cite{Nam:2009nn} and Refs.~\cite{Harrington:1976dj,Diakonov:1988my} to this end. Usually, there are two different instanton configurations at finite $T$, being periodic in Euclidean time, with trivial and nontrivial holonomies. They are called the Harrington-Shepard~\cite{Harrington:1976dj} and Kraan-Baal-Lee-Lu calorons~\cite{Kraan:1998pm,Lee:1998bb}, respectively. The nontrivial holonomy can be identified as the Polyakov line as an order parameter for the confinment-deconfinement transition of QCD. However, since we are not interested in the transition in this work as long as we are at relatively low $T$, we choose the Harrington-Shepard caloron for the modifications at finite $T$. Note that here are two caveats: 1) These modifications are done for a pure glue system without dynamical quarks. Hence, the instanton variables may change, if one takes into account dynamical-quark contributions in the instanton distribution function. 2) Moreover, we assume a $CP$-invariant vacuum for the modifications, i.e.  $\Delta=0$, whereas we are interested in the physical quantities for $\Delta\ne0$ for CME. Correcting these inconsistencies may give rise to changes in the final results, although they seem small considering that the order of $CP$ violation in reality is very tiny, but it must be beyond our scope in the present work. Keeping this issue in mind, we write the instanton distribution function at finite $T$ with the Harrington-Shepard caloron as follows:
%EQUATION>>>
\begin{equation}
\label{eq:IND}
d(\rho,T)=\underbrace{C_{N_c}\,\Lambda^b_{\mathrm{RS}}\,
\hat{\beta}^{N_c}}_\mathcal{C}\,\rho^{b-5}
\exp\left[-(A_{N_c}T^2
+\bar{\beta}\gamma n\bar{\rho}^2)\rho^2 \right].
\end{equation}
%EQUATION<<<
Here, the abbreviated notations are also given as:
%EQUATION>>>
\begin{equation}
\label{eq:para}
\hat{\beta}=-b\ln[\Lambda_\mathrm{RS}\rho_\mathrm{cut}],\,\,\,\,
\bar{\beta}=-b\ln[\Lambda_\mathrm{RS}\langle R\rangle],\,\,\,
C_{N_c}=\frac{4.60\,e^{-1.68\alpha_{\mathrm{RS}} Nc}}{\pi^2(N_c-2)!(N_c-1)!},
\end{equation}
%EQUATION<<<
%EQUATION>>>
\begin{equation}
\label{eq:AA}
A_{N_c}=\frac{1}{3}\left[\frac{11}{6}N_c-1\right]\pi^2,\,\,\,\,
\gamma=\frac{27}{4}\left[\frac{N_c}{N^2_c-1}\right]\pi^2,\,\,\,\,
b=\frac{11N_c-2N_f}{3},\,\,\,\,n=\frac{N}{V}.
\end{equation}
%EQUATION<<<
Note that we defined the one-loop inverse charge $\hat{\beta}$ and $\bar{\beta}$ at a certain phenomenological cutoff value $\rho_\mathrm{cut}$ and $\langle R\rangle\approx\bar{R}$. As will be shown, only $\bar{\beta}$ is relevant in the following discussions and will be fixed self-consistently within the present framework. $\Lambda_{\mathrm{RS}}$ stands for a scale, depending on a renormalization scheme, whereas $V_3$ stands for the three-dimensional volume. Using the instanton distribution function in Eq.~(\ref{eq:IND}), we can compute the average value of the instanton size, $\bar{\rho}^2$ straightforwardly as follows~\cite{Schafer:1996wv}:
%EQUATION>>>
\begin{equation}
\label{eq:rho}
\bar{\rho}^2(T)
=\frac{\int d\rho\,\rho^2 d(\rho,T)}{\int d\rho\,d(\rho,T)}
=\frac{\left[A^2_{N_c}T^4
+4\nu\bar{\beta}\gamma n \right]^{\frac{1}{2}}
-A_{N_c}T^2}{2\bar{\beta}\gamma n},
\end{equation}
%EQUATION<<<
where $\nu=(b-4)/2$. Substituting Eq.~(\ref{eq:rho}) into Eq.~(\ref{eq:IND}), the distribution function can be evaluated further as:
%EQUATION>>>
\begin{equation}
\label{eq:dT}
d(\rho,T)=\mathcal{C}\,\rho^{b-5}
\exp\left[-\mathcal{M}(T)\rho^2 \right],\,\,\,\,
\mathcal{M}(T)=\frac{1}{2}A_{N_c}T^2+\left[\frac{1}{4}A^2_{N_c}T^4
+\nu\bar{\beta}\gamma n \right]^{\frac{1}{2}}.
\end{equation}
%EQUATION<<<
The instanton-number density $n$ can be computed self-consistently as a function of $T$, using the following equation:
%EQUATION>>>
\begin{equation}
\label{eq:NOVV}
n^\frac{1}{\nu}\mathcal{M}(T)=\left[\mathcal{C}\,\Gamma(\nu) \right]^\frac{1}{\nu},
\end{equation}
%EQUATION<<<
where we have replaced $NT/V_3\to n$, and $\Gamma(\nu)$ indicates a $\Gamma$ fucntion with an argument $\nu$. Note that $\mathcal{C}$ and $\bar{\beta}$ can be determined easily using Eqs.~(\ref{eq:rho}) and (\ref{eq:NOVV}), incorporating the vacuum values of the $n$ and $\bar{\rho}$: $\mathcal{C}\approx9.81\times10^{-4}$ and $\bar{\beta}\approx9.19$. At the same time, using these results, we can obtain the average instanton size $\bar{\rho}$ as a function of $T$ with Eq.~(\ref{eq:rho}).

\begin{figure}[t]
\begin{tabular}{cc}
\includegraphics[width=8.5cm]{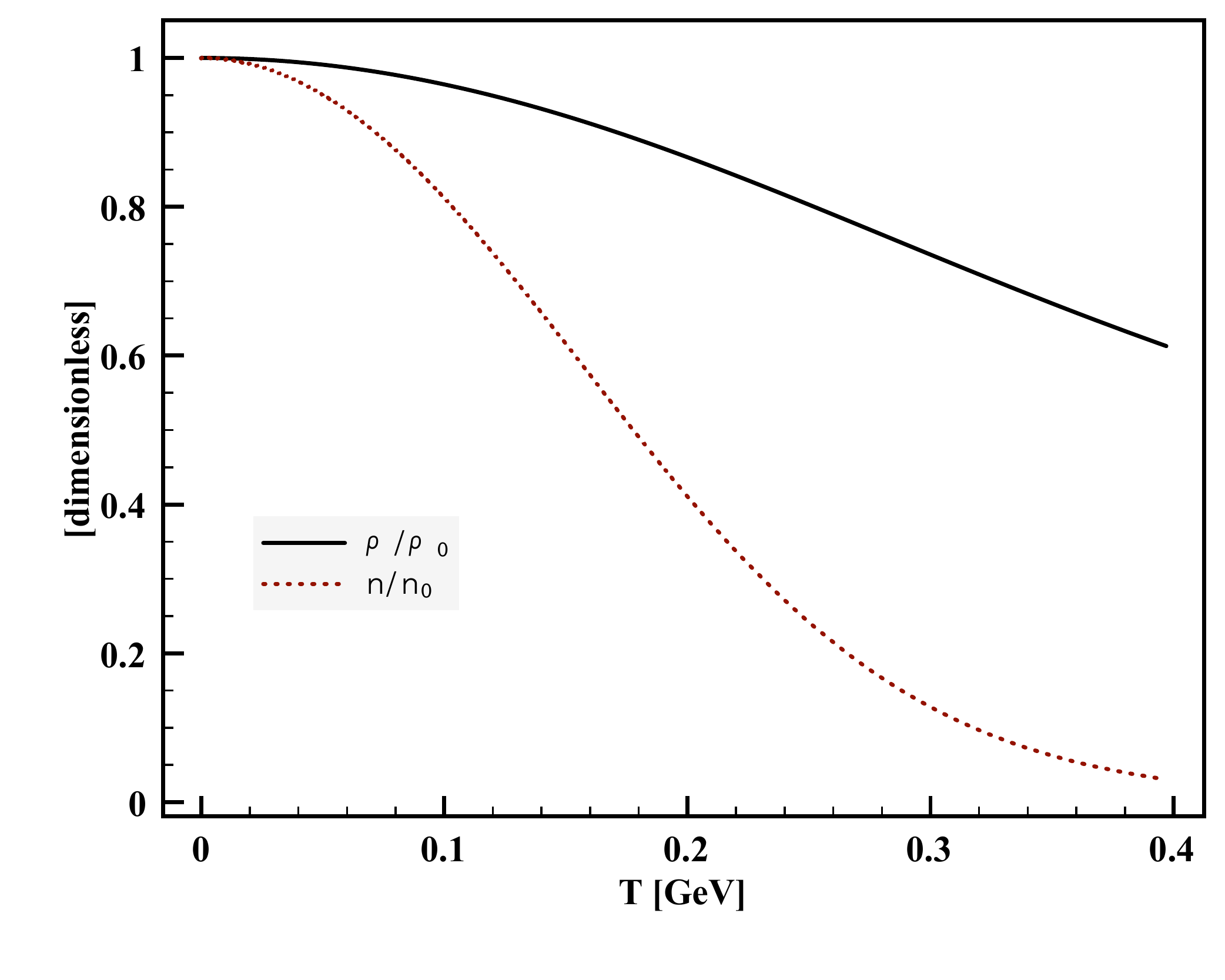}
\includegraphics[width=9.5cm]{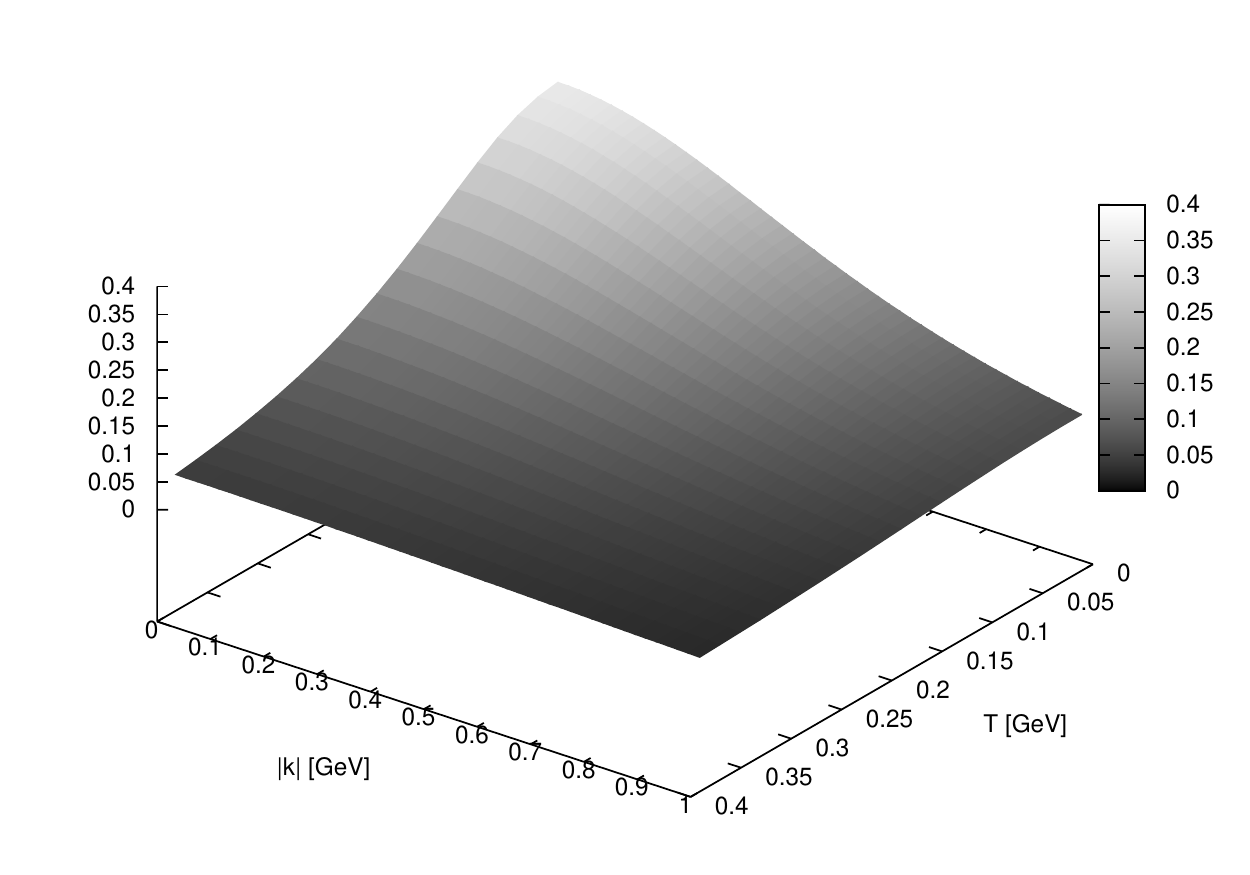}
\end{tabular}
\caption{(Color online) Normalized $\bar{\rho}/\bar{\rho}_{0}$ and $n/n_{0}$ as a function of $T$ for $N_{c}=3$ (left). $M$ as a function of $T$ and absolute value of the momentum $|\bm{k}|$ (right).}       
\label{FIG1}
\end{figure}

Finally, in order to estimate the $T$ dependence of the constituent-quark mass $M_{0}$, it is necessary to consider the normalized distribution function, defined as follows:
%EQUATION>>>
\begin{equation}
\label{eq:NID}
d_N(\rho,T)=\frac{d(\rho,T)}{\int d\rho\,d(\rho,T)}
=\frac{\rho^{b-5}\mathcal{M}^\nu(T)
\exp\left[-\mathcal{M}(T)\rho^2 \right]}{\Gamma(\nu)}.
\end{equation}
%EQUATION<<<
Now, we want to employ the large-$N_c$ limit to simplify the expression of $d_N(\rho,T)$. Since the parameter $b$ is in the order of $\mathcal{O}(N_c)$ as shown in Eq.~(\ref{eq:para}), it becomes infinity as $N_c\to\infty$, and the same is true for $\nu$. In this limit, as understood from Eq.~(\ref{eq:NID}), $d_N(\rho,T)$ can be approximated as a $\delta$ function~\cite{Diakonov:1995qy}: 
%EQUATION>>>
\begin{equation}
\label{eq:NID2}
\lim_{N_c\to\infty}d_N(\rho,T)=\delta[{\rho-\bar{\rho}\,(T)}].
\end{equation}
%EQUATION<<<
Considering the constituent-quark mass can be represented by~\cite{Diakonov:1995qy} 
%EQUATION>>>
\begin{equation}
\label{eq:M0}
M_{0}\propto\sqrt{n}\int d\rho\,\rho^{2}\delta[\rho-\bar{\rho}(T)]
=\sqrt{n(T)}\,\bar{\rho}^{2}(T),
\end{equation}
%EQUAITON<<<
we can modify $M_{0}$ as a function of $T$ as follows:
%EQUATION>>>
\begin{equation}
\label{eq:momo}
M_{0}\to M_{0}\left[\frac{\sqrt{n(T)}\,\bar{\rho}^2(T)}
{\sqrt{n(0)}\,\bar{\rho}^2(0)}\right]\equiv M_{0}(T)
\end{equation}
%EQUATION<<<
where we will use $M_{0}\approx325$ MeV as done for zero $T$. The numerical results for the normalized $\bar{\rho}/\bar{\rho}_{0}$ and $n/n_{0}$ as functions of $T$ are in the left panel of Fig.~\ref{FIG1}. As shown there, these quantities are decreasing with respect to $T$ as expected: decreasing instanton effect. However, even beyond $T^{\chi}_{c}\approx\Lambda_{\mathrm{QCD}}\approx200$ MeV, the instanton contribution remains finite. In the right panel of figure, we draw the quark mass as a function of $T$ and absolute value of three momentum of a quark $|\bm{k}|$:
%EQUATION>>>
\begin{equation}
\label{eq:M00}
M(|\bm{k}|,T)=M_{0}(T)\left[\frac{2}{2+\bar{\rho}^{2}(T)\,|\bm{k}|^{2}}\right].
\end{equation}
%EQUAITON<<<
Note that we have ignored the Euclidean-time component of the four momentum by setting $k_{4}=0$. This tricky treatment simplifies the calculations in hand to a large extent, and we also verified that only a small deviation appears in comparison to full calculations. Moreover, $\bar{\rho}$ in Eq.~(\ref{eq:M00}) is now a function of $T$ as demonstrated by Eqs.~(\ref{eq:rho}) and (\ref{eq:momo}) previously. As shown in the figure, $M(|\bm{k}|,T)$ is a smoothly decreasing function of $T$ and $|\bm{k}|$, indicating that the effect of the instanton is diminished.  For more details, one can refer to the previous work~\cite{Nam:2009nn}.
%FIGURE>>>
%FIGURE<<<
%----------------------------------------------------------------------------------------------------
\section{Vector-current correlation at finite temperature}
%----------------------------------------------------------------------------------------------------
In this Section, we briefly discuss how connected VCC in Eq.~(\ref{eq:SPEC}) are converted as a function of $T$. For this purpose, we make use of the fermionic Matsubara formula (Euclidean-time formula) as follows:
%EQUATION>>>
\begin{equation}
\label{eq:MATS}
\int\frac{d^4k}{(2\pi)^4}F(k)\to
T\sum_{n=-\infty}^{\infty}\int\frac{d^3\bm{k}}{(2\pi)^3}F(\bm{k},w_{n}),
\end{equation}
%EQUAITON<<<
where the fermionic Matsubara frequency reads $w_{n}=(2n+1)\pi T$. Using this formula, the relevant four vectors in Eq.~(\ref{eq:FF}) can be rewritten by
%EQUATION>>>
\begin{equation}
\label{eq:MO}
k^{2}=w^{2}_{n}+E^{2}_{a},\,\,\,\,(k-q)^{2}=w^{2}_{n}+E^{2}_{b}.
\end{equation}
%EQUAITON<<<
where $k=(0,\mathrm{k}\sin\theta,\mathrm{k}\cos\theta,w_{n})$ with $\mathrm{k}\equiv|\bm{k}|$ and $q=(0,0,0,|Q|)$ in the space-like region for the corresponding vector meson at rest for simplicity. The energies of the quarks involved are also defined by
%EQUATION>>>
\begin{equation}
\label{eq:EEEE}
E^{2}_{a}=\mathrm{k}^{2}+(1+\delta^{2})M^{2}_{a},\,\,\,\,
E^{2}_{b}=\mathrm{k}^{2}-2w_{n}|Q|+|Q|^{2}
+(1+\delta^{2})M^{2}_{b}\approx\mathrm{k}^{2}-2w_{0}|Q|+|Q|^{2}
+(1+\delta^{2})M^{2}_{b}.
\end{equation}
%EQUAITON<<<
Note that we have kept the term $2w_{n}|Q|$ for $E_{b}$ and assumed that the lowest Matsubara frequency $w_{0}=\pi T$ is dominant. Later, we will see that to keep this term is important to have correct $|Q|$ behaviors of VCC. Similarly, the $T$-and momentum-dependent quark masses in Eq.~(\ref{eq:MF}) are also redefined as follows:
%EQUATION>>>
\begin{equation}
\label{eq:MMMM}
M_{a}=M_{0}\left(\frac{2}{2+\mathrm{k}^{2}\bar{\rho}^{2}} \right)^{2},\,\,\,\,
M_{b}=M_{0}\left[\frac{2}{2+(\mathrm{k}^{2}-2w_{0}|Q|+|Q|^{2})\bar{\rho}^{2}} \right]^{2}.
\end{equation}
%EQUAITON<<<
The $T$ dependence of these masses are included in $M_{0}$ as shown in the previous Section. As understood in Eq.~(\ref{eq:MF}), $\bar{M}_{a,b}$ and $\tilde{M}_{a,b}$ are also similarly redefined. We write a useful summation identity which will be used to evaluate the denominator in the expressions for VCC:
%EQUATION>>>
\begin{eqnarray}
\label{eq:SUM}
&&\sum_{n=-\infty}^{\infty}\left[\frac{T}{(w^{2}_{n}+X^{2})
(w^{2}_{n}+Y^{2})^{2}} \right]
=\frac{1}{8TXY^{3}(X^{2}-Y^{2})^{2}}\mathrm{sech}^{2}\left(\frac{Y}{2T}\right)
\cr
&&\times
\left[XY^{3}-YX^{3}+4TY^{3}\mathrm{cosh}^{2}\left(\frac{Y}{2T}\right)
\mathrm{tanh}\left(\frac{X}{2T}\right)
+TX(X^{2}-3Y^{2})\mathrm{sinh}\left(\frac{Y}{T}\right) \right]
\equiv\mathcal{M}(X,Y).
\end{eqnarray}
%EQUAITON<<<

Considering all the ingredients discussed so far, finally, we arrive at expressions for the imaginary parts of connected VCC as a function of $T$, strength of the external magnetic field $B_{0}$, and strength of $P$ and $CP$ violations $\delta$:
%EQUATION>>>
\begin{eqnarray}
\label{eq:PI1234}
\mathrm{Im}\Pi_{11,22}(q)
=-2|Q|\delta B_{0}N_{c}N_{f}
\int\frac{\mathrm{k}^{3}d\mathrm{k}\,\cos\theta d\cos\theta}{4\pi^2}
\left[M_{a}\bar{M}_{a}\mathcal{M}(E_{b},E_{a})
- M_{b}\bar{M}_{b}\mathcal{M}(E_{a},E_{b}) \right],
\cr
\mathrm{Im}\Pi_{33}(q)
=-2|Q|\delta B_{0}N_{c}N_{f}
\int\frac{\mathrm{k}^{3}d\mathrm{k}\,\cos\theta d\cos\theta}{4\pi^2}
\left[M_{a}\bar{M}_{a}\mathcal{M}(E_{b},E_{a})
+ M_{b}\bar{M}_{b}\mathcal{M}(E_{a},E_{b}) \right],
\cr
\mathrm{Im}\Pi_{44}(q)
=-4|Q|\delta B_{0}N_{c}N_{f}
\int\frac{\mathrm{k}^{3}d\mathrm{k}\,\cos\theta d\cos\theta}{4\pi^2}
\left[M_{a}\bar{M}_{a}\mathcal{M}(E_{b},E_{a})
+M_{b}\bar{M}_{b}\mathcal{M}(E_{a},E_{b}) \right],
\end{eqnarray}
%EQUAITON<<<
where we have used Eqs.~(\ref{eq:F1234}), (\ref{eq:SPEC}), and (\ref{eq:SUM}), and the isospin indices are chosen to be $a=b$ for simplicity. As mentioned previously, the longitudinal and temporal components become different to each other, according to the breakdown of Lorentz invariance in the temporal direction at finite $T$.

%--------------------------------------------------------------------------------------------------
\section{Charge separation, induced magnetic field, and centrality in HIC}
%--------------------------------------------------------------------------------------------------
In this Section, we want to discuss ChS defined in Eq.~(\ref{eq:CHA}) using the results given in the previous Section and the previous work ~\cite{Nam:2009jb,Nam:2009hq}. Considering the fact that $|\langle J_{\perp}\rangle_{B_{0},\delta}/\langle J_{\parallel}\rangle_{B_{0},\delta}| \propto\delta\ll1$ and $\delta\ll1$ shown in Ref.~\cite{Nam:2009jb}, we can ignore the term $\langle J_{\perp}\rangle^{2}_{B_{0},\delta}$ rather safely in Eq.~(\ref{eq:CHA}). Moreover, as will be shown shortly, $\lim_{|Q|\to0}\mathrm{Im}\Pi_{11,22}\propto\chi_{(1,2),B_{0},\delta}$ becomes almost zero due to the negative sign between the terms inside the square bracket in the last  line of Eq.~(\ref{eq:PI1234}). Hence, we can drop the term $\chi_{\perp,B_{0},\delta}$, resulting in the following expression for ChA: 
%EQUAITON<<<
\begin{equation}
\label{eq:RET}
\langle\langle \cos(\Delta\phi_{a}+\Delta\phi_{b})\rangle\rangle
\propto\langle J_{3}\rangle^{2}_{B_{0},\delta}
+\chi_{3,B_{0},\delta}.
\end{equation}
%EQUAITON<<<
Referring to the previous works~\cite{Nam:2009jb,Nam:2009hq}, $\langle J_{3}\rangle^{2}_{B_{0},\delta}$ reads:
%EQUATION>>>
\begin{equation}
\label{eq:JJ}
\langle J_{3}\rangle^{2}_{B_{0},\delta}=4\mu^{2}_{\chi}B^{2}_{0}
V^{2}_{3}N^{2}_{c}
\left[\int\frac{\mathrm{k}^{2}d\mathrm{k}}
{2\pi^2}M_{a}\bar{M}_{a}\bar{\mathcal{M}}(E_{a}) \right]^{2},
\end{equation}
%EQUAITON<<<
where we have used the chiral-chemical potential $\mu^{2}_{\chi}=\delta^{2}A^{2}_{0}$~\cite{Nam:2009jb}. Note that we employed $A_{0}$, being instead of $iA_{4}$, which stands for fluctuations in the external-static EM field as in Eq.~(\ref{eq:A}). We also included the three-dimensional volume $V_{3}$, resulting in that Eq.~(\ref{eq:JJ}) becomes dimensionless~\cite{Fukushima:2008xe}. The relevant function in Eq.~(\ref{eq:JJ}) coming from the Matsubara sum is written as follows:
%EQUATION>>>
\begin{equation}
\label{eq:MMM}
\bar{\mathcal{M}}(E_{a})=
\frac{1}{E^{2}_{a}}\left[\frac{1}{E_{a}}
\frac{(1-e^{-E_{a}/T})}{(1+e^{-E_{a}/T})} 
-\frac{1}{2T}\frac{e^{-E_{a}/T}}{(1+e^{-E_{a}/T})^{2}}\right].
\end{equation}
%EQUAITON<<<
As mentioned, VCC in Eq.~(\ref{eq:PI1234}) in the limit of $|Q|\to0$ can be related with the background contribution or the susceptibility:
%EQUATION>>>
\begin{equation}
\label{eq:BBBB}
\chi_{3,B_{0},\delta}\equiv \lim_{|Q|\to0}
\left[\frac{V_{3}T}{N_{f}}\mathrm{Im}\Pi_{33}(|Q|) \right].
\end{equation}
%EQUAITON<<<
Here we took an average over the flavor. Again, Eq.~(\ref{eq:BBBB}) is dimensionless. More explicitly, Eq.~(\ref{eq:BBBB}) can be evaluated further and resulted in:
%EQUATION>>>
\begin{eqnarray}
\label{eq:JJJ}
\chi_{3,B_{0},\delta}&=&
-2\mu_{\chi} B_{0}V_{3}N_{c}T\lim_{|Q|\to0}
\left[
\int\frac{\mathrm{k}^{3}d\mathrm{k}\,\cos\theta d\cos\theta}{4\pi^2}
\left[M_{a}\bar{M}_{a}\mathcal{M}(E_{b},E_{a})
+ M_{b}\bar{M}_{b}\mathcal{M}(E_{a},E_{b}) \right] \right]
\cr
&=&-4\mu_{\chi} B_{0}V_{3}N_{c}T
\int\frac{\mathrm{k}^{3}d\mathrm{k}}{2\pi^2}
M_{a}\bar{M}_{a}\mathcal{M}(E_{a},E_{a}),
\end{eqnarray}
%EQUAITON<<<
where
%EQUATION>>>
\begin{equation}
\label{eq:BMBM}
\mathcal{M}(E_{a},E_{a})=
\frac{1}{32T^{2}E^{5}_{a}}
\left\{6T^{2}\mathrm{tanh}\left(\frac{E_{a}}{2T} \right)
-E_{a}\mathrm{sech}^{2}\left(\frac{E_{a}}{2T} \right)
\left[3T+E_{a}\mathrm{tanh}\left(\frac{E_{a}}{2T} \right) \right]
 \right\}.
\end{equation}
%EQUAITON<<<
In deriving Eq.~(\ref{eq:JJJ}), we have assumed the following: As seen in Eq.~(\ref{eq:PI1234}), the longitudinal component of connected VCC contains a term $|Q|\delta$. Considering the relation between $\delta$ and the chiral-chemical potential as above, we can write $|Q|\delta=\mu_{\chi}|Q|/A_{0}$. Since the fluctuation $A_{0}$ must be far smaller than the model scale $\sim600$ MeV and can be considered nearly zero, we assume the following relation in the limit of $(|Q|,A_{0})\to0$:
%EQUATION>>>
\begin{equation}
\label{eq:RATIO}
\lim_{|Q|,A_{0}\to0}\left(\frac{|Q|}{A_{0}}  \right)\sim1.
\end{equation}
%EQUAITON<<<
Due to this simple assumption, one is lead to $|Q|\delta\sim\mu_{\chi}$. As a result of all the ingredients discussed so far, the relevant terms for ChS, $\langle J_{3}\rangle^{2}_{B_{0},\delta}$ and $\chi_{3,B_{0},\delta}$ are expressed as functions of $B_{0}$, $\mu_{\chi}$, and $T$ in Eqs.~(\ref{eq:JJ}) and (\ref{eq:JJJ}).  

\begin{figure}[t]
\includegraphics[width=8.5cm]{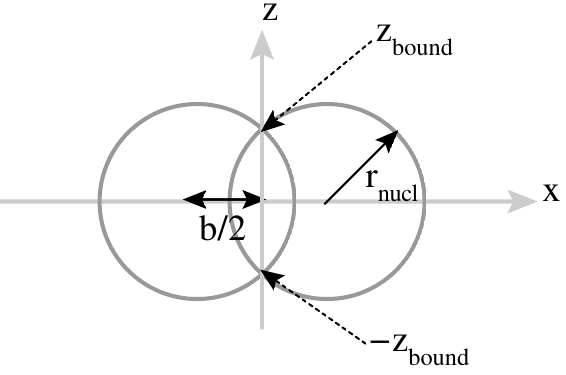}
\caption{Schematic figure for the transverse plane of a hevay-ion collision. $r_{\mathrm{nucl}}$ and $b/2$ indicate the radius of the nucleus and the distance between the nucleus center and the $z$ axis. The magnetic field is generated along the $z$ axis. Thus, the case with $b=0$ stands for a head-on collision, whereas $b=2r_{\mathrm{nucl}}$ indicates the case without collision.}       
\label{FIG00}
\end{figure}

Now, we are in a position to have a geometric consideration of HIC to derive the strength of the magnetic field as a function of centrality $(or N_{\mathrm{part}})$, so as to estimate ChS. A schematic figure for a simple geometry of HIC on the transverse plane is depicted in Fig.~\ref{FIG00}. Note that, since we have set the magnetic field along the $z$ axis, the transverse plane is so defined by the $x$-$z$ plane, being different from the usual definition exploited in HIC, in which the $x$-$y$ plane is chosen for it. As shown in the figure, the radius of the nucleus is assigned as $r_{\mathrm{nucl}}$, while the normal distance between the center of the nucleus and the imaginary center line of the collision as $b/2$, in which $b$ denotes the impact parameter. For later convenience, we define the region of the overlap of the two nuclei as {\it overlap region}. In Ref.~\cite{Kharzeev:2007jp}, the magnetic field induced along HIC was given by the Li\'enard-Wiechert potential at $(\bm{r}-\bm{r}_{0})$, where $\bm{r}_{0}$ indicates the position vector of the charged particle on the transverse plane:
%EQUATION>>>
\begin{equation}
\label{eq:LW}
\bm{B}(\bm{r})
=\frac{Z\,\alpha_{\mathrm{EM}}\,\mathrm{sinh}Y\,
[(\bm{r}-\bm{r}_{0})\times\hat{y}]}
{\left[(\bm{r}-\bm{r}_{0})^{2}
+(t\,\mathrm{sinh}Y-y\,\mathrm{cosh}Y)^{2} \right]^{3/2}},
\end{equation}
%EQUAITON<<<
where $Z$, $\alpha_{\mathrm{EM}}$, and $Y$ denote the electric charge of the nucleus, fine-structure constant, and rapidity, respectively. If we assume only the very early stage of the collision, $t\approx y\approx0$, the magnetic field reside on the transverse plane with a small rapidity $Y\ll1$. For instance, the beam rapidity $Y_{\mathrm{beam}}$ for the Au+Au collision is about $5.371$ for $\sqrt{s}=200$ GeV per a nucleon pair in the nucleus, using the following relation
%EQUATION>>>
\begin{equation}
\label{eq:RA}
Y=\frac{1}{2}\ln\left[\frac{p_{0}+p_{y}}{p_{0}-p_{y}} \right],
\end{equation}
%EQUAITON<<<
where $p_{0}$ and $p_{y}$ is the energy and momentum along the beam ($y$) direction. Then the rapidity $Y$ can locate at $-Y_{\mathrm{beam}}\le Y\le Y_{\mathrm{beam}}$. Considering the above assumption, the magnetic field can be simplified as follows: 
%EQUATION>>>
\begin{equation}
\label{eq:LW11}
B_{0}(z)\approx\frac{Z\,\alpha_{\mathrm{EM}}\,Y\,b}
{[z^{2}+(b/2)^{2}]^{3/2}},
\end{equation}
%EQUAITON<<<
where we have chosen $\bm{r}=\bm{z}$ and $\bm{r}_{0}=(b/2)\hat{x}$, since we focus on the magnetic field induced on the $z$ axis and assumed that the distribution of the magnetic field along the $x$ axis is small. It is worth mentioning that there are two electric-charge sources for the magnetic field in HIC: those from the participants inside the overlap region and spectators outside region~\cite{Kharzeev:2007jp}. However, at the very early stage of the collision, the contributions from the two charge sources can be nearly the same. In other words, $\mathrm{sinh}Y$ varies slowly. Moreover, similarly, the electric-charge sources are averagely locating at $x=\pm b/2$ for the early stage as in the assumption. Now, we define a useful quantity, which is assigned as:
%EQUATION>>>
\begin{equation}
\label{eq:ZBOUND}
z_{\mathrm{bound}}=\sqrt{r^{2}_{\mathrm{nucl}}-(b/2)^{2}}.
\end{equation}
%EQUAITON<<<
As understood by seeing Fig.~\ref{FIG00}, this lengthy quantity bounds the $z$ value of the overlap region. As discussed in Ref.~\cite{Kharzeev:2007jp}, the magnetic field is screened so that ChS develops mostly from the surface of the overlap region. At the  same time, this screening effect makes ChS for SCC is larger than that of OCC as discussed in Ref.~\cite{Kharzeev:2007jp}. Hence the screening effect can be parameterized by
%EQUATION>>>
\begin{eqnarray}
\label{eq:FFFFSSS}
F_{\mathrm{scr},\pm\pm}&=&\exp\left[-2\alpha_{\mathrm{scr}}|z_{\mathrm{bound}}-z| \right]
+\exp\left[-2\alpha_{\mathrm{scr}}|z_{\mathrm{bound}}+z|\right]
\cr
F_{\mathrm{scr},\pm\mp}&=&2\exp\left[-\alpha_{\mathrm{scr}}|z_{\mathrm{bound}}-z|\right]
\exp\left[-\alpha_{\mathrm{scr}}|z_{\mathrm{bound}}+z| \right],
\end{eqnarray}
%EQUAITON<<<
where $\alpha_{\mathrm{scr}}$ indicates a parameter in GeV, which relates to the screening length of QGP: $\alpha_{\mathrm{scr}}\propto1/\lambda_{\mathrm{QGP}}$. Considering  all the ingredients discussed so far, we can write ChS for SCC and OCC in the following expressions~\cite{Kharzeev:2007jp}: 
%EQUATION>>>
\begin{eqnarray}
\label{eq:ChSSSS}
\langle\langle \cos(\Delta\phi_{\pm}+\Delta\phi_{\pm})\rangle\rangle
&\propto&\mathcal{C}_{\pm\pm}
\int^{ z_{\mathrm{bound}}}_{- z_{\mathrm{bound}}}
\frac{dz}{2\pi}\left(\langle J_{3}\rangle^{2}_{B_{0},\delta}+\chi_{3,B_{0},\delta} \right)F_{\mathrm{scr},\pm\pm},
\cr
\langle\langle \cos(\Delta\phi_{\pm}+\Delta\phi_{\mp})\rangle\rangle
&\propto&\mathcal{C}_{\pm\mp}
\int^{ z_{\mathrm{bound}}}_{- z_{\mathrm{bound}}}
\frac{dz}{2\pi}\left(\langle J_{3}\rangle^{2}_{B_{0},\delta}+\chi_{3,B_{0},\delta} \right)F_{\mathrm{scr},\pm\mp}.
\end{eqnarray}
%EQUAITON<<<
Note that the magnetic field and screening effect are integrated over the overlap region along the $z$ axis. Here $\mathcal{C}_{ab}$ is a coefficient as a function of the positive- and/or negative-charge particles observed in HIC experiments. For instance, it reads~\cite{Fukushima:2009ft}:
%EQUATION>>>
\begin{equation}
\label{eq:CVFVC}
\mathcal{C}_{ab}=\frac{ab}{N_{a}N_{b}}.
\end{equation}
%EQUAITON<<<
To make a problem easy, we set $\mathcal{C}_{\pm\pm}\approx-\mathcal{C}_{\pm\mp}$ for brevity hereafter.  The three-dimensional volume appearing inside $\langle J_{3}\rangle^{2}_{B_{0},\delta}$ and $\chi_{3,B_{0},\delta}$ in Eqs.~(\ref{eq:JJ}) and (\ref{eq:BBBB}) can be estimated for the overlap region as follows:
%EQUATION>>>
\begin{equation}
\label{eq:VOL}
V_{3}\approx\pi z_{\mathrm{bound}} \frac{b}{2} \bar{r}_{\mathrm{nucl}}
=\gamma\pi z_{\mathrm{bound}} \frac{b}{2}r_{\mathrm{nucl}}.
\end{equation}
%EQUAITON<<<
Note that $V_{3}$ above is a disc with the overlap area $\pi  z_{\mathrm{bound}}b/2$ and height $\bar{r}_{\mathrm{nucl}}$, which is a Lorentz-contracted radius of the nucleus in the $y$ direction, i.e. $\gamma r_{\mathrm{nucl}}$. The Lorentz-contraction factor $\gamma$ becomes about 100 for $\sqrt{s}=200$ GeV. Now we want to discuss the domain with  $Q_{t}\ne0$ for CME produced in HIC. Since the size of the domain does not depend on specific centralities for collisions~\cite{Kharzeev:2007jp,:2009txa,:2009uh}, the probability for finding the domain in a certain volume of QGP created in HIC can be regarded proportional to the size of the nucleus, which may corresponds to multiplicity, inversely. Similarly, the probability for that the magnetic field induced at the domain can be the same. Therefore, the volume in Eq.~(\ref{eq:VOL}) and the magnetic field in Eq.~(\ref{eq:LW11}) are assumed to be modified as 
%EQUATION>>>
\begin{equation}
\label{eq:VB}
V_{3}\to \frac{V_{3}}{N_{\mathrm{nucl}}},\,\,\,\,
B_{0}\to \frac{B_{0}}{N_{\mathrm{nucl}}}.
\end{equation}
%EQUAITON<<<
In this way, we put the effect of multiplicity for the different nuclei collisions to ChS. Later, we will see that these modifications play an important role to reproduce the correct strength hierarchy of ChS for different kinds of nuclei depending on their size and atomic mass.

The next step is to define centrality with $N_{\mathrm{part}}$, which is represented by the number of the participants in HIC, as a function of $b$. Since we already have an expression for the induced magnetic field as a function of $b$ as well as $z$ in Eq.~(\ref{eq:LW11}), by doing this, we can relate the magnetic field with centrality of the collision. Again, accounting for the collision geometry given in Fig.~\ref{FIG00}, the centrality $N_{\mathrm{part}}$ can be 
parameterized with $b$ as follows:
%EQUATION>>>
\begin{equation}
\label{eq:CCC}
N_{\mathrm{part}}=2N_{\mathrm{nucl}}
\left\{1-\frac{2}{\pi}
\left[
\mathrm{tan}^{-1}\left(\frac{b}
{2\sqrt{r^{2}_{\mathrm{nucl}}-(b/2)^{2}}} \right)
+\frac{b}{2r^{2}_{\mathrm{nucl}}}
\sqrt{r^{2}_{\mathrm{nucl}}-(b/2)^{2}}\right] \right\},
\end{equation}
%EQUAITON<<<
where $N_{\mathrm{nucl}}$ is the atomic mass of the projectile nucleus in HIC. It is easily shown that $N_{\mathrm{part}}=2N_{\mathrm{nucl}}$ for the head-on collision ($b=0$) and $N_{\mathrm{part}}=0$ for the case without a collision ($b/2=r_{\mathrm{nucl}}$). For instance, the radius of a gold nucleus is about $r_{\mathrm{nucl}}\approx7.27$ fm with $N_{\mathrm{Au}}\approx197$ simply using the well-known relation $r_{\mathrm{nucl}}=1.25\,\mathrm{fm}\times N^{3}_{\mathrm{nucl}}$. Thus, $b/2$ can vary from $0$ to about $7.27$ fm for the Au+Au collision. Then,  numerical results for ChS will be given as a function of centrality defined as
%EQUATION>>>
\begin{equation}
\label{eq:CCCEN}
\mathrm{Centrality}\equiv\frac{(2N_{\mathrm{Nucl}}-N_{\mathrm{part}})}
{2N_{\mathrm{nucl}}}\times100.
\end{equation}
%EQUAITON<<<

%FIGURE>>>
%FIGURE<<<
%-------------------------------------------------
\section{Numerical results and Discussions}
\begin{figure}[t]
\begin{tabular}{cc}
\includegraphics[width=8.5cm]{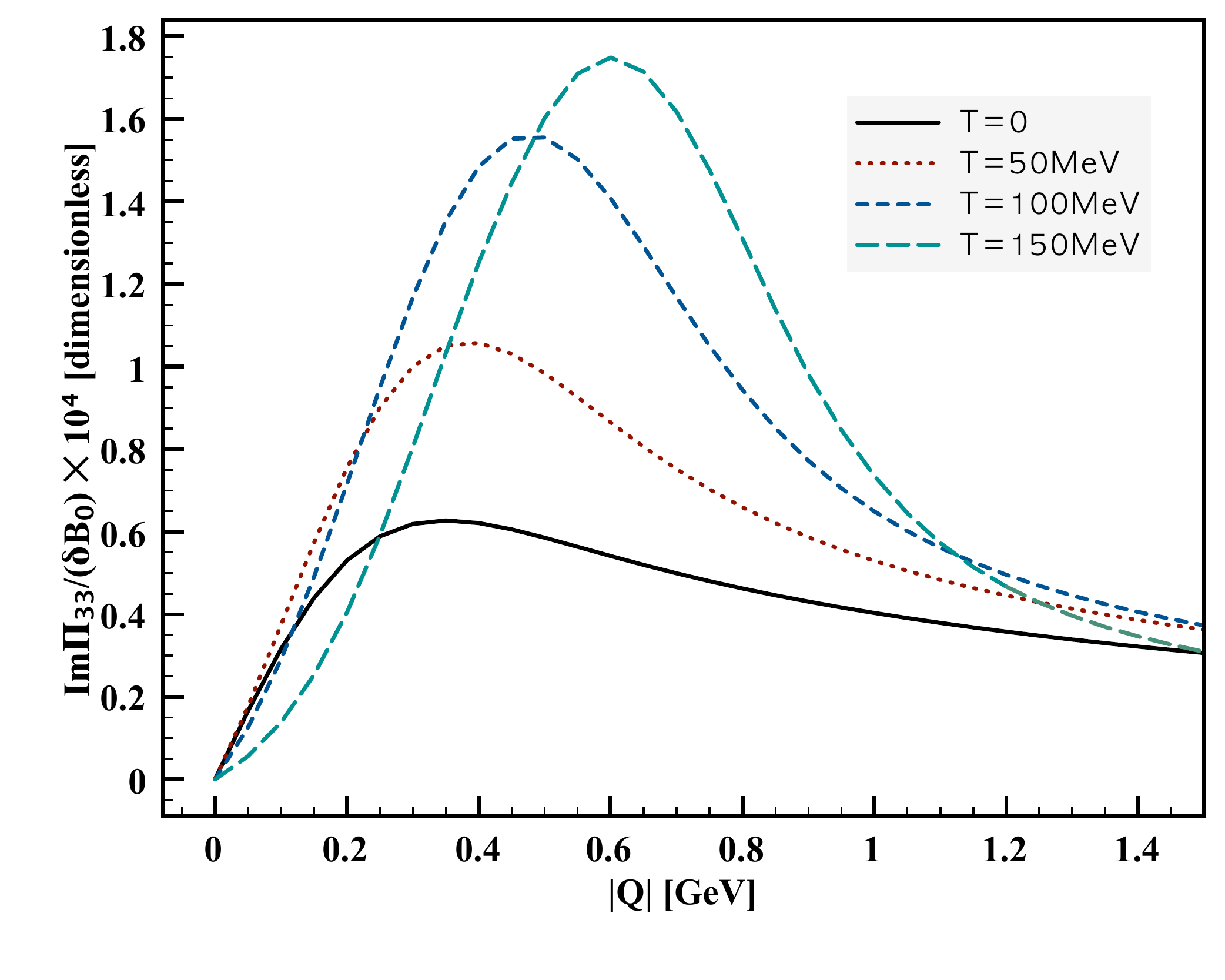}
\includegraphics[width=8.5cm]{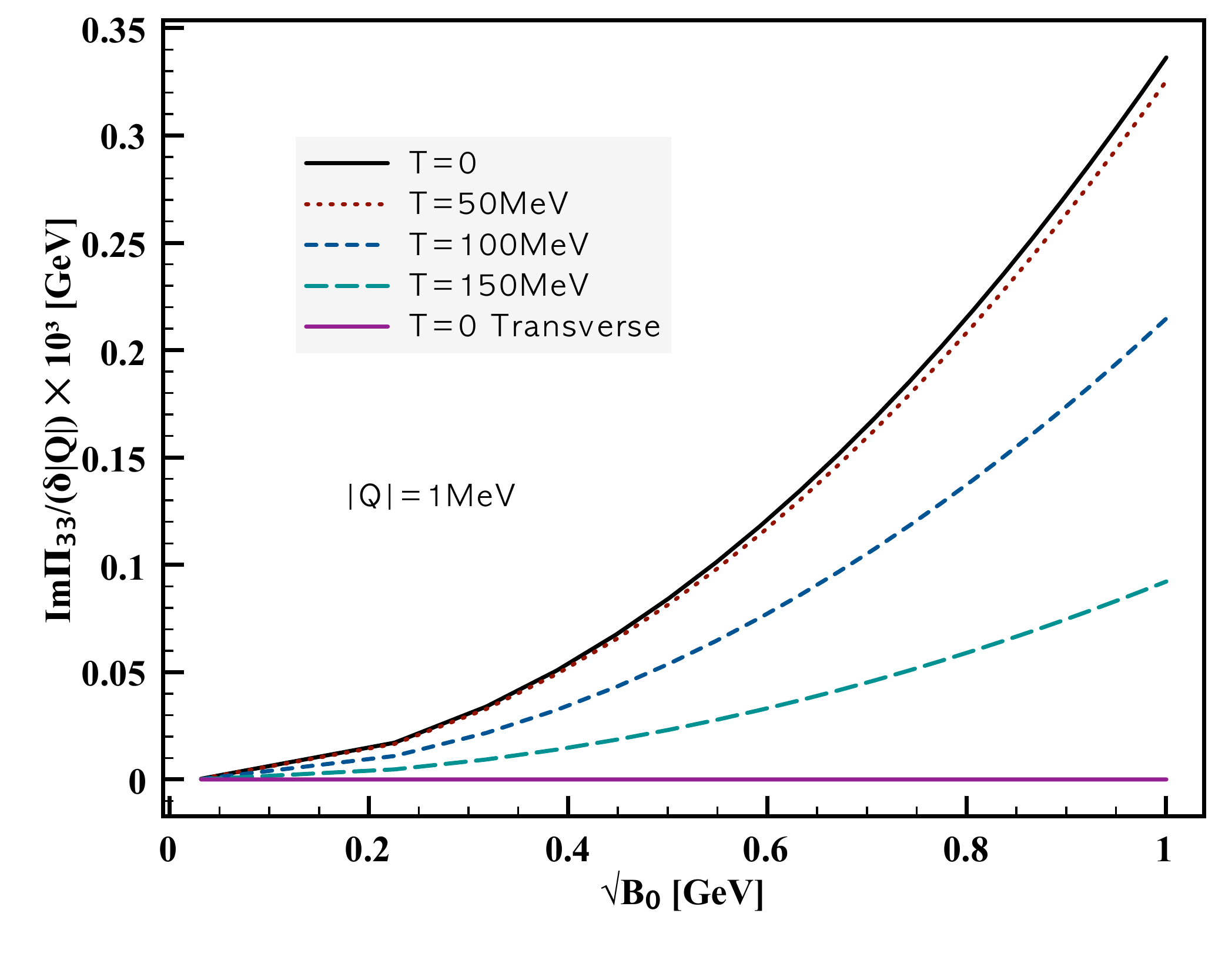}
\end{tabular}
\caption{(Color online) $\mathrm{Im}\Pi_{33}/(\delta B_{0})$ as a function of $|Q|$ (left) and $\mathrm{Im}\Pi_{33}/(\delta |Q|)$ as a function of $\sqrt{B_{0}}$  at $|Q|=1$ MeV (right) for different $T$. In the right panel, we also draw $\mathrm{Im}\Pi_{11,22}/(\delta |Q|)$ for $T=0$ in a thick-solid line.}       
\label{FIG2}
\end{figure}

%-------------------------------------------------
In this Section, we provide numerical results for the physical quantities discussed in the previous Sections. First, we consider connected VCC defined in Eq.~(\ref{eq:PI1234}). In the left panel of Fig.~\ref{FIG2}, we show the numerical results for $\mathrm{Im}\Pi_{33}$ as a function of $|Q|$ for different $T=(0,50,100,150)$ MeV. For simplicity, we divided it with $\delta$ and $B_{0}$, since VCC is linear in these two quantities as in Eq.~(\ref{eq:PI1234}). Thus, one can obtain it for different values of $\delta$ and $B_{0}$ by a simple scaling. 

As in the figure, at $T=0$, we obtain a smooth curve, which has a wide bump in the vicinity of $|Q|=(300\sim400)$ MeV, which presents a vector meson corresponding to the current in Eq.~(\ref{eq:VC}). Since we have picked up only the terms proportional to $\mathcal{O}(B_{0})$ and $\mathcal{O}(\delta)$ to examine CME in the present work, the bump position does not match with possible physical vector-meson mass, such as $\rho(770)$ for example. As $T$ increases, the magnitude of  the curve becomes larger and the bump position shifts to a higher value of $|Q|$. Moreover, the bump becomes more narrowed. For instance, we observe a bump at $|Q|\approx600$ MeV for $T=150$ MeV. We note that this enhancing behavior of the magnitude depends much on the lowest frequency $w_{0}$ included in $E_{b}$ as well as $M_{b}$. This indicates thermal modification of the corresponding vector-meson mass. In Ref.~\cite{Buividovich:2010tn}, VCC and corresponding vector-meson spectral function were obtained using a SU(2) quenched lattice QCD simulation under a strong magnetic field, and their results for the vector-meson spectral function indeed showed the same tendency with that shown in the left panel of Fig.~\ref{FIG2}. On the contrary, the bump position in their work locates at about $1$ GeV for $T=0$, which is higher than ours and may correspond to the mass of $\rho$ meson. Although we did not show the transverse component of VCC, it behaves similar to that of the longitudinal ones but in different magnitudes. 

Here is some physical interpretations on the behaviors shown in the figure. We already mentioned that it depends much on the Matsubara frequency inside $E_{b}$ and $M_{b}$. In addition, as noted previously and  in Ref.~\cite{Diakonov:1988my,Nam:2009nn}, the instanton contribution gets diminished as $T$ increases, being indicated by decreasing average instanton size and increasing average inter-instanton distance shown in the right panel of Fig.~\ref{FIG1}, i.e. diluting instanton medium.  As understood in Eq.~(\ref{eq:F1234}), the relevant part of quark propagator for understanding the problem can be simply written as follows:
%EQUATION>>>
\begin{equation}
\label{eq:CCCCC}
\frac{1}{(k-q)^{2}+M^{2}}\approx
\frac{1}{k^{2}-(2\pi T)|Q|+|Q|^{2}+M^{2}},
\end{equation}
%EQUAITON<<<
where we have approximated $k_{4}$ with the lowest Matsubara frequency $w_{0}=\pi T$. In the light-hand side of the above equation, it can be easily noticed that the term $(2\pi T)|Q|$ plays a kind of a thermal mass, at which a bump is produced, while the term $M^{2}$ gets smaller due to diluting instanton. Moreover the term $(2\pi T)|Q|$ gets larger linearly with respect to $T$, being consistent with the observation. We verified that, if we choose a higher Matsubara frequency, the bump position appears at higher $|Q|$ values, as expected. For instance, we observe a bump at $400$ MeV with $w_{0}$ and at $600$ MeV with $w_{1}$ for $T=50$ MeV. At the same time, the magnitude of VCC become enhanced for higher $T$, since the denominator of Eq.~(\ref{eq:CCCCC}) gets smaller, being consistent with that of the lattice QCD simulation~\cite{Buividovich:2010tn}. The narrowing bump with respect to $T$ also can be explained by decreasing instanton contribution, resulting in smaller $M$ in Eq.~(\ref{eq:CCCCC}).

In the right panel of Fig.~\ref{FIG2}, we show VCC in the limit of $|Q|\to0$ as a function of $\sqrt{B_{0}}$. Note that this quantity is proportional to the susceptibility $\chi_{3,B_{0},\delta}$ as in Eq.~(\ref{eq:BBBB}). In the calculation, we took $|Q|=1$ MeV as a small value to see the behaviors of VCC in the limit of $|Q|\to0$ for numerical convenience. Otherwise, one can use Eq.~(\ref{eq:JJJ}) for correct limiting behaviors. As shown in the figure, the longitudinal component turns out to be finite and increases as the magnetic field gets stronger as easily understood by Eq.~(\ref{eq:PI1234}). In contrast, the transverse one uniformly zero due to the  cancelation between the two terms inside the square bracket in the right-hand side of the first line of Eq.~(\ref{eq:PI1234}), since $M_{a}$ equals $M_{b}$ and, subsequently, $\bar{M}$ and $\mathcal{M}$ are the same for $q=0$, equivalently $|Q|=0$. In other words, the transverse component at $\mathcal{O}(B_{0})$ must be zero for all the values of $B_{0}$ at $|Q|\to0$.

Again, this observation is well consistent with that of Ref.~\cite{Buividovich:2010tn}, in which VCC is represented by the corresponding conductivities $\sigma_{zz}$ and $\sigma_{xx}$. However, there appear differences between theirs and the present calculation as $T$ goes higher. The lattice result showed almost flat but nonzero values for the longitudinal and transverse components of VCC for higher $T$. We interpret this difference by that there can be thermal enhancement of VCC for $\mathcal{O}(B^{n}_{0})$ for $n\ne1$ which has not been included in the present work. In our calculations, as for the longitudinal component of VCC, one finds that it becomes insensitive to $B_{0}$ or, equivalently, the slope of the curve gets smaller with respect to $B_{0}$ for higher $T$, being consistent with the previous work~\cite{Nam:2009jb,Nam:2009hq}. This behavior can be understood as follows: Since CME is an effect depending on the topological behaviors of the nontrivial QCD vacuum. Hence, if the vacuum contribution decreases due to some reasons, CME also does. We already emphasized that the instanton contribution, which represents the nontrivial vacuum, gets weaker and diluted as $T$ increases. As a result, CME and corresponding VCC under the strong magnetic field becomes weak with respect to $T$. In this sense, although we do not go beyond $T^{\chi}_{c}$ in the present work, considering the tendency shown in the figure, it is easy to imagine that the slope of the longitudinal component of VCC goes to zero for considerably high $T$, then becomes flat for $B_{0}$ finally. A consistent situation was observed in the lattice simulation, but with a finite magnitude~\cite{Buividovich:2010tn}. This finiteness (nonzero), being different from ours, again can be explained by the thermal enhancement
%FIGURE>>>
%FIGURE<<<

Now, we are in a position to compute ChS from the instanton contribution at low $T$ defined in Eq.~(\ref{eq:ChSSSS}) using the present and previous~\cite{Nam:2009jb} results for connected and disconnected VCC. Then, ChS can be written Explicitly as follows:
%EQUATION>>>
\begin{equation}
\label{eq:DFDF}
\langle\langle \cos(\Delta\phi_{a}+\Delta\phi_{b})\rangle\rangle
\propto2\mathcal{C}_{ab}
\int
\frac{dz}{\pi}\left\{\bar{\mathcal{C}}^{2}
\left[\int\frac{\mathrm{k}^{2}d\mathrm{k}}
{2\pi^2}M_{a}\bar{M}_{a}\bar{\mathcal{M}}(E_{a}) \right]^{2}
-\bar{\mathcal{C}}T
\int\frac{\mathrm{k}^{3}d\mathrm{k}}{2\pi^2}
M_{a}\bar{M}_{a}\mathcal{M}(E_{a},E_{a}) \right\}
F_{\mathrm{scr},ab},
\end{equation}
%EQUAITON<<<
where we have simply ignored the transverse contributions of VCC as already mentioned and defined a coefficient $\bar{\mathcal{C}}$ as a function of $z$ for convenience:
%EQUATION>>>
\begin{equation}
\label{eq: }
\bar{\mathcal{C}}(z)\equiv
\frac{\mu_{\chi} B_{0}(z)V_{3}N_{c}}{N^{2}_{\mathrm{nucl}}}.
\end{equation}
%EQUAITON<<<
Since we have no available experimental information on $\mu_{\chi}$, we set it as $\mu_{\chi}=1$ MeV as a trial, considering that $P$- and $CP$-violation strengths are small. Moreover, the value of $\mathcal{C}_{ab}$ is chosen as $-1$ and $+1$ for SCC and OCC, respectively, due to that ChS can be easily scaled for realistic numbers of $N_{a,b}$ as in Eq.~(\ref{eq:DFDF}). Note that we use $|\mathcal{C}_{ab}|=1$ even for different nuclei collisions, since the dependence of multiplicity ($\sim$ size of the nuclei) has been already included in the magnetic field and volume as in Eq.~(\ref{eq:VB}).

\begin{figure}[t]
\includegraphics[width=8.5cm]{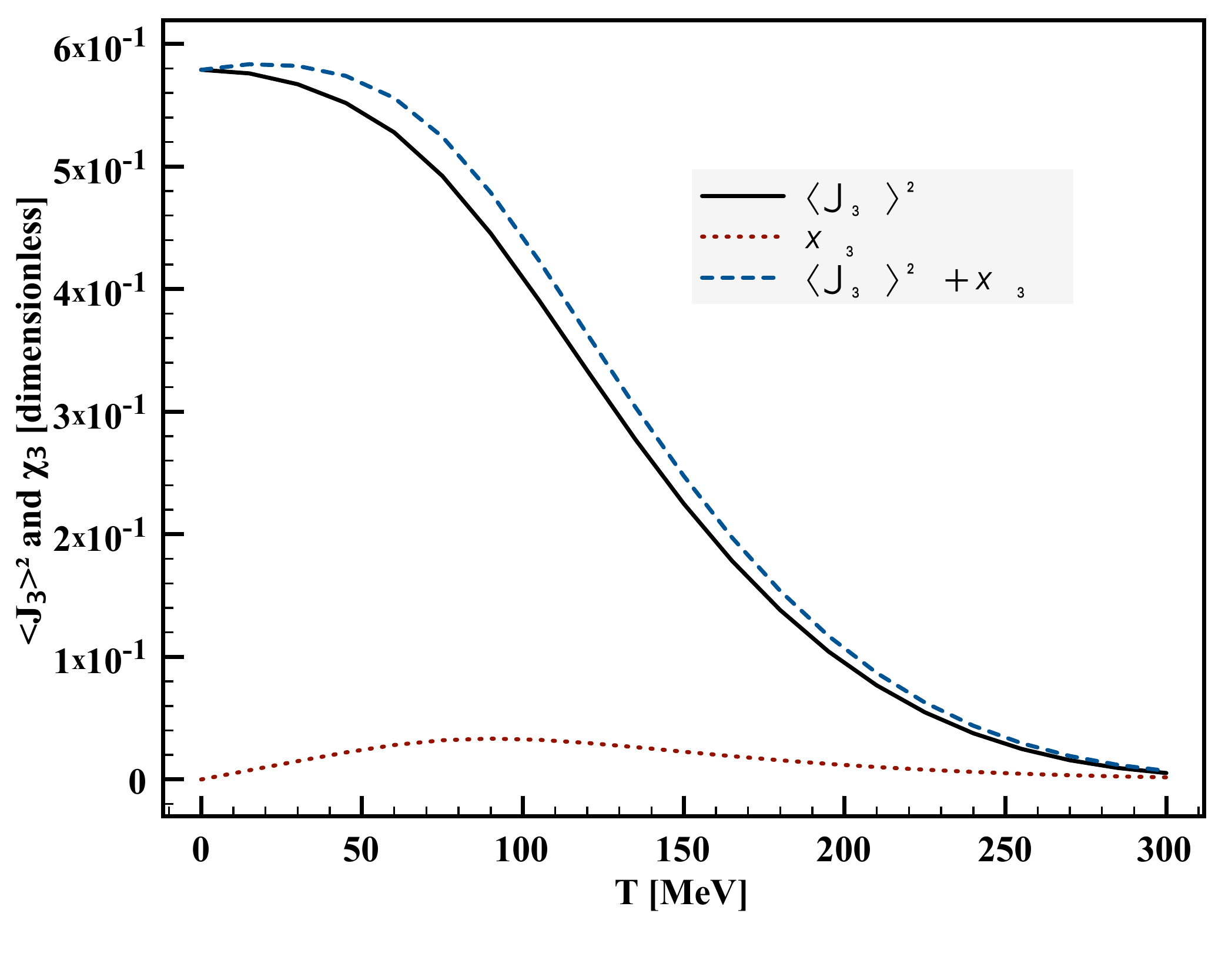}
\caption{(Color online) $\langle J_{3}\rangle^{2}$, $\chi_{3}$, and  $\langle J_{3}\rangle^{2}+\chi_{3}$ in Eqs.~(\ref{eq:JJ}) and (\ref{eq:JJJ}) as functions of $T$. Here we use $B_{0}=m^{2}_{\pi}$ and $V_{3}=4\pi r^{3}_{\mathrm{Au}}/3$. }       
\label{FIG3}
\end{figure}

Before going further, we first look at the $T$-dependence behavior of $\langle J_{3}\rangle^{2}$ and $\chi$ in Eqs.~(\ref{eq:JJ}) and (\ref{eq:JJJ}). Here, we have dropped the subscripts $B_{0}$ and $\delta$ from the quantities for simplicity. In Fig.~\ref{FIG3}, we depict them as functions of $T$ up to $3$ GeV, which is obviously beyond the applicability of the present framework though. In computing them, we set $B_{0}=m^{2}_{\pi}$ and $V_{3}=4\pi r^{2}_{\mathrm{Au}}/3$ as a trial. As shown in the figure, $\langle J_{3}\rangle $ shows a decreasing curve with respect to $T$ as already observed in the previous work~\cite{Nam:2009jb}: Diluting instanton for higher $T$. As for $\chi$, we find a small bump in the vicinity $T=(50\sim100)$ MeV, then it decreases. Obviously, the magnitude of $\chi$ is much larger than that of $\langle J_{3}\rangle $ over the region, whereas the difference between them is diminished and alsmost disappears  at $T\approx300$ MeV. Since the sum of these two quantities is responsible for ChS as shown in Eq.~(\ref{eq:DFDF}), we also show the sum of them in the figure. As expected, their contribution to ChS is dominated by $\langle J_{3}\rangle^{2}$. This observation is consistent with that given in Ref.~\cite{Fukushima:2009ft}. Hence, taking into account for the previous and present results, the instanton contribution for ChS (or CME) can be almost inferred from the longitudinal disconnected VCC, $\langle J_{3}\rangle^{2}$ alone. 
%FIGURE>>>
%FIGURE<<<

\begin{figure}[t]
\begin{tabular}{cc}
\includegraphics[width=8.5cm]{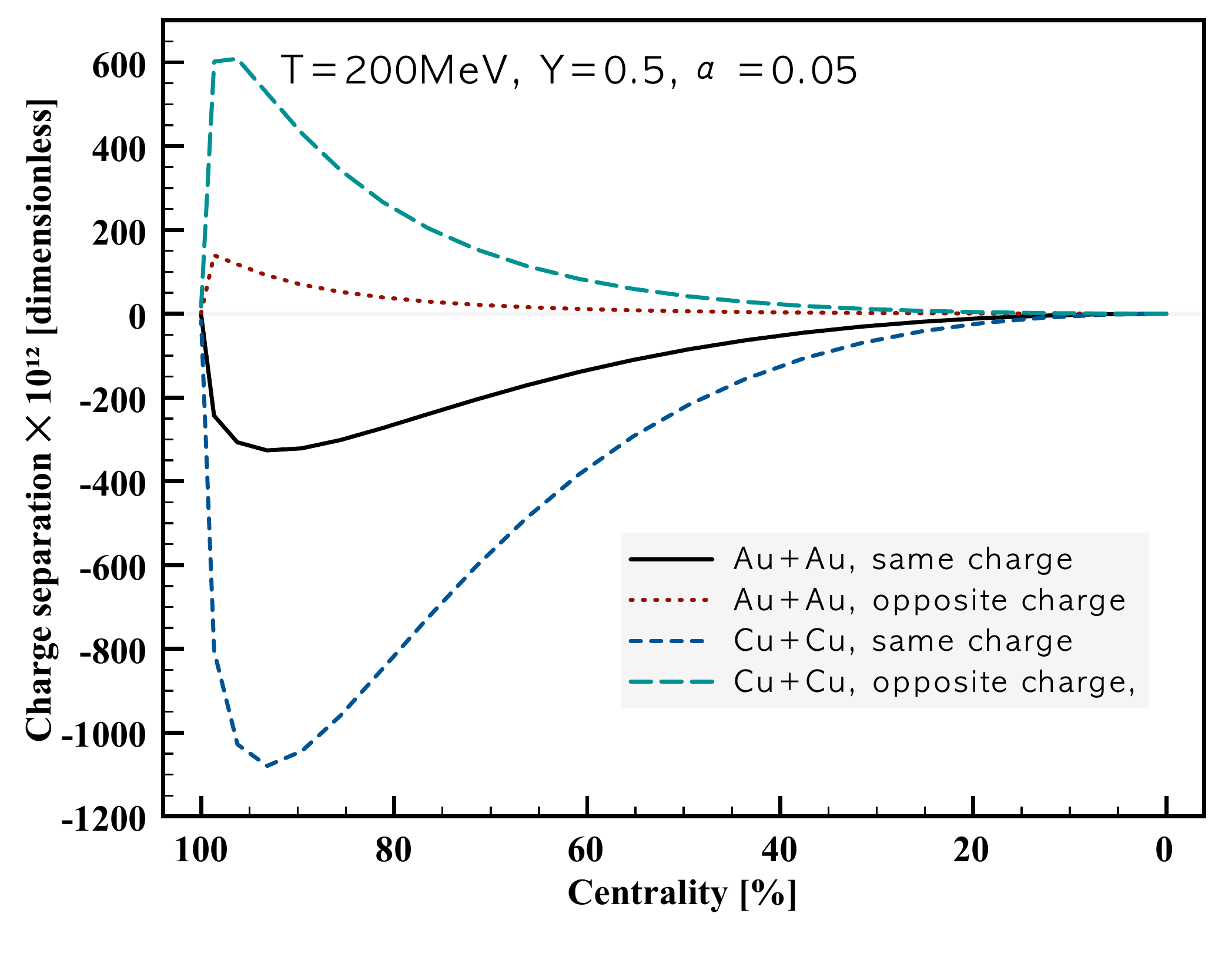}
\includegraphics[width=8.5cm]{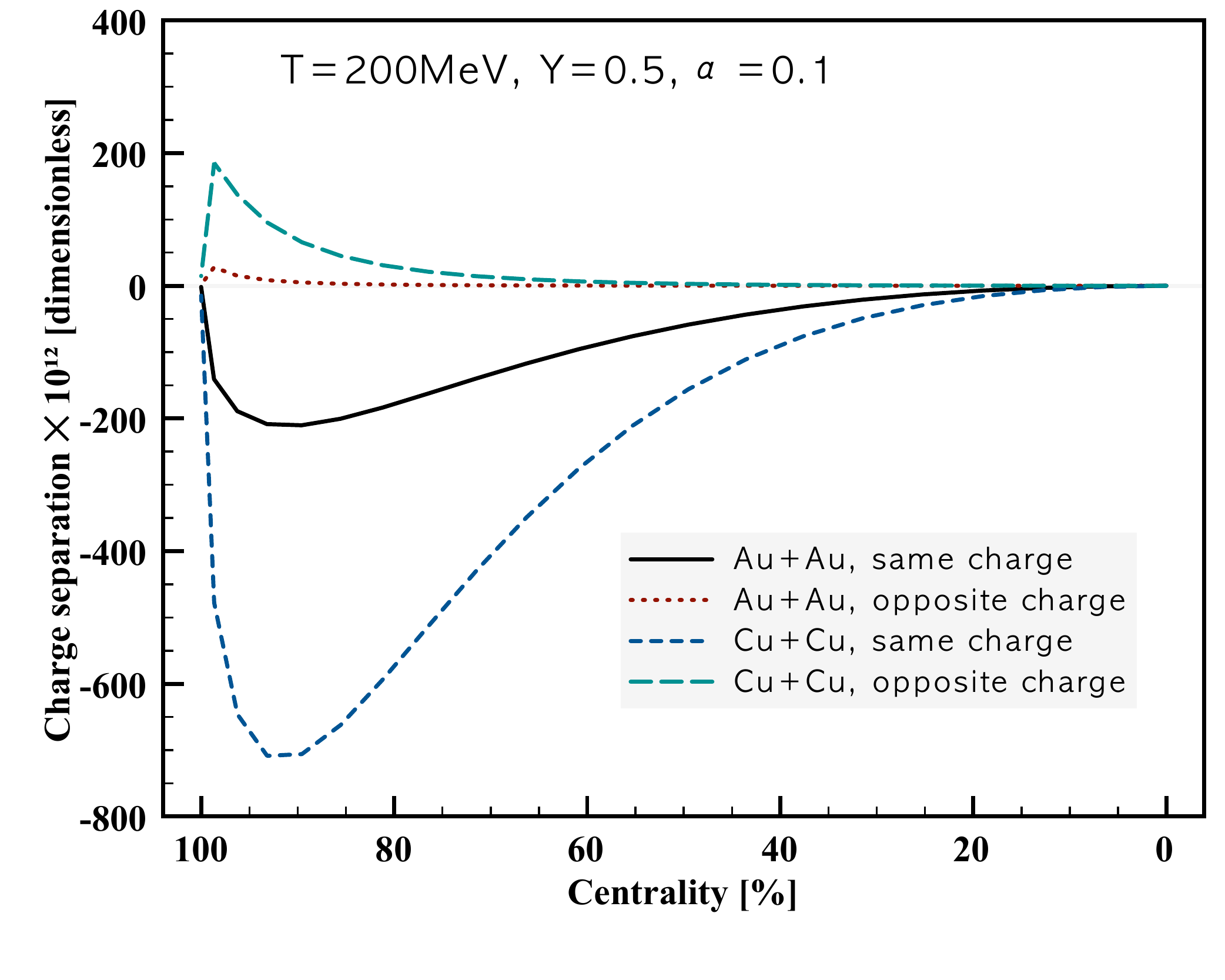}
\end{tabular}
\caption{(Color online) Charge separation in Eq.~(\ref{eq:DFDF}) as a function of centrality for the Au+Au and Cu+Cu collisions for $\alpha=0.05$ (left) and $0.1$ (right). We choose $T=200$ MeV and $Y=0.5$.}       
\label{FIG4}
\end{figure}

In Fig.~\ref{FIG4}, we draw the curves for ChS defined in Eq.~(\ref{eq:DFDF})  as a function of centrality in Eq.~(\ref{eq:CCCEN}). We show separately SCC and OCC for ChS for the Au+Au and Cu+Cu collisions for $T=200$ MeV and $Y=0.5$. The curves in the left panel are plotted with $\alpha_{\mathrm{scr}}=0.01$ GeV, whereas the right one with $\alpha_{\mathrm{scr}}=0.5$ GeV. Although we did not show the numerical reuslt for different $T$, as for lower $T$ (higher $T$), one can easily expect that the curves will increase (decrease) according to the $T$ dependence of the instanton contributions shown Fig.~\ref{FIG3}. First, we take a look on the left panel of the figure. In general, it turns out that the strengths of ChS (or CME) gets increasing from head-on to peripheral collisions. The reason for this increasing is that the induced magnetic field by the collision becomes stronger as $b$ increases as understood in Eq.~(\ref{eq:LW11}). However, as the two nuclei passing by each other without a collision, in other words $b\to2r_{\mathrm{nucl}}$, ChS decreases drastically, since the overlap region disappears by Eq.~(\ref{eq:VOL}), i.e. $z_{\mathrm{bound}}=0$. 

As for the Au+Au collision, we observe that SCC turns out to be considerably larger than OCC shown in the left panel. This behavior can be understood by the screening effect given by Eq.~(\ref{eq:FFFFSSS}): If a particle drifting by CME moves inward to the overlap region, the probability to find it after hadronization is much less than that for a particle outward from the region~\cite{Kharzeev:2007jp}. In this sense, the screening effect controls the difference between the strengths for SCC and OCC. We observe a similar tendency also for the Cu+Cu collision as seen in the figure. However, we find a distinctive differences from those for the Au+Au collision: 1) the magnitudes of the curves for SCC and OCC get larger and 2) the difference between SCC and OCC is diminished in comparison to those for the Au+Au collision. As mentioned previously, the magnetic field and volume have been modified, accounting for that the domain for $Q_{t}\ne0$ is not depending on centrality, with the probability $\sim1/N_{\mathrm{nucl}}$. Due to this modification, there is strong suppressions for ChS for the heavier (or larger) nuclei collision. This is the reason for 1). In contrast, the reason for 2) is a combing consequence of the modification and screening effect. In other words, the smaller difference in SCC  and OCC means the weak screening effect. Since the radius for Cu is smaller than that for Au, the probability that a particle via CME drifts through the overlap region then becomes hadronized gets higher. This tendency is clearly demonstrated by comparing the left ($\alpha_{\mathrm{scr}}=0.05$) and right ($\alpha_{\mathrm{scr}}=0.1$) panels of the figure. We verified that, as we take smaller values for $\alpha_{\mathrm{nucl}}$, the difference between SCC and OCC becomes smaller, and vice versa for the larger values.

Here we make a brief discussion on the the value of $\alpha_{\mathrm{scr}}$. In Ref.~\cite{Kharzeev:2007jp}, it was proposed that $\alpha_{\mathrm{scr}}=1/\lambda_{\mathrm{QGP}}$. Considering the screening length can be estimated as $\lambda_{\mathrm{QGP}}\propto1/(gT)$, where $g$ stands for the strong coupling, at $T=200$ MeV, the length is in the order of less than one fermi. If we take $\alpha_{\mathrm{scr}}=1/\lambda_{\mathrm{QGP}}$ with $\alpha_{\mathrm{scr}}=0.05$ GeV and $0.1$ GeV, the screening lengths become about $4$ fm and $2$ fm, respectively. Thus, these values are relatively larger. However, since there must be more unknown factors in HIC than what have done in the present work, the simple parametrization for the screening effect in Eq.~(\ref{eq:FFFFSSS}) does not compete fully with the experiment, and one needs a tuning for the parameters. Moreover, the collision geometry was oversimplified here so that the larger $\alpha_{\mathrm{nucl}}$ may be necessary to compensate what have ignored in the present work. 

We note that the simple collision geometry taken into account in the present work is also valid for different sources of CME, such as the sphaleron, as long as the CME current is a linear function of the external magnetic field. Considering that the sphaleron contribution is also linear in $B_{0}$ as in Refs.~\cite{Kharzeev:2007jp,Fukushima:2008xe}, the behaviors of the curves in Fig.~\ref{FIG4} can be valid for higher $T$ but in different magnitudes. In this sense, if we compare the present results for ChS with to the experimental data from STAR collaboration~\cite{:2009txa,:2009uh}, the curves shown in the left panel of the figure may be consistent qualitatively with the data, showing the correct strength hierarchy for SCC and OCC for the different types of collisions. Moreover, the shape of the curves are similar to the data.

%FIGURE>>>
%FIGURE<<<

%-------------------------------------------------
\section{Summary and conclusion}
%-------------------------------------------------
In the present work, we have investigated VCC and ChS, which was suggested as an indication of CME. To this end, we employed the instanton-vacuum configuration with the $T$ modification using the Harrington-Shepard caloron. All the relevant quantities were presented as functions of $\delta$, which amounts the strength of $P$- and $CP$ violations. The external magnetic field was included by the linear Schwinger method. Using all these ingredients, we obtained expressions for connected and disconnected VCC. The imaginary parts of connected VCC relating to a spectral function for a vector meson as well as conductivity were computed. Considering a simple collision geometry of HIC and using the present results for VCC, we calculated ChS as a functions of centrality. Below, we summarize the important observations of the present work briefly:
%ITEMIZE>>>
\begin{itemize}
\item In leading contributions, connected VCC turns out to be linear in $B_{0}$ as well as $\delta$.
\item The imaginary part of connected VCC shows a wide bump at $|Q|=(300\sim400)$ MeV for $T=0$. The bump structure is enhanced and moves toward a higher $|Q|$ value with respect to $T$, due to the increasing thermal-mass-like effect. At the same time, the bump becomes sharpened by decreasing instanton contribution. 
\item Taking the limit $|Q|\to0$, the transverse component of connected VCC disappears by the cancelation between relevant terms, whereas the longitudinal one remains finite. As $T$ increases, VCC becomes insensitive to $B_{0}$, denoting a decreasing CME due to diluting instanton. 
\item Considering a simple collision geometry for HIC, and assuming the magnetic field generated in a very early stage of the collision and screening effect, we can estimate ChS for the Au+Au and Cu+Cu collisions. The strength of ChS for SCC turns out to be larger than that for OCC, due to the screening effect, in accordance with experiments. 
\item ChS for the lighter nucleus collision is more enhanced than that for the heavier one, since CME is proportional to $1/N^{2}_{\mathrm{nucl}}$. It also turns out that the difference between SCC and OCC gets smaller as the nucleus involved becomes lighter. However, this tendency largely depends on the strength of the screening effect.  
\item This simple collision geometry estimation on ChS is almost independent on which source of CME plays a role, instanton or sphaleron, as far as the induced CME current is linear in $\mu_{\chi}B_{0}$ as in Eq.~(\ref{eq:GJ}). Therefore, the present estimation for ChS can be compared with actual HIC experiments to a certain extent.
\end{itemize}
%ITEMIZE>>>

We note that there were several simplifications and assumptions to be addressed more carefully in the present work. For  instance, we have only picked up the terms proportional to $\mathcal{O}(B_{0})$ and $\mathcal{O}(\delta)$ for VCC. By construction, we cannot describe the confinement-deconfinement transition, resulting in that we are reluctant to go beyond the critical $T$ for S$\chi$SB, $T^{\chi}_{c}\approx\Lambda_{\mathrm{QCD}}$. Moreover, the collision geometry was oversimplified which leads to insufficient considerations on the complicated macroscopic behaviors of the QGP matter. We do not determine the screening parameter from the present model, and  it is treated as a free parameter. Especially,  the time evolution of  ChS was totally ignored. However, all the results given in the present work may describe the essence of what happens for VCC as well as ChS in HIC, and provide information on the relatively low-$T$ aspect of CME. In addition, the simple collision geometry will provide an almost model-independent estimation on HIC, although it is still far from full descriptions on the experiments. More sophisticated treatment for the collision geometry, parametrization of $B_{0}$, screening effect, and time evolution are under progress, and appears elsewhere.   
%-------------------------------------------------
\section*{Acknowledgment}
%-------------------------------------------------
The author thanks C.~W.~Kao and J.~W.~Chen for fruitful discussions. He is also grateful to K.~S.~Choi for technical supports on the numerical calculations. This work was supported by the grant NSC 98-2811-M-033-008 from National Science Council (NSC) of Taiwan. All the figures were generated using JaxoDraw (http://jaxodraw.sourceforge.net/), Plot (http://plot.micw.eu/), and Gnuplot (http://www.gnuplot.info/).
%--------------------------------------------------


\begin{thebibliography}{99}
%--------------------------------------------------
\bibitem{Hirano:2002ds}
  T.~Hirano and K.~Tsuda,
  %``Collective flow and two pion correlations from a relativistic  hydrodynamic
  %model with early chemical freeze out,''
  Phys.\ Rev.\  C {\bf 66}, 054905 (2002).
%  [arXiv:nucl-th/0205043].
%--------------------------------------------------
\bibitem{Heinz:2001xi}
  U.~W.~Heinz and P.~F.~Kolb,
  %``Early thermalization at RHIC,''
  Nucl.\ Phys.\  A {\bf 702}, 269 (2002)
%  [arXiv:hep-ph/0111075].
%--------------------------------------------------
\bibitem{Kharzeev:2004ey}
  D.~Kharzeev,
  %``Parity violation in hot QCD: Why it can happen, and how to look for it,''
  Phys.\ Lett.\  B {\bf 633}, 260 (2006).
%  [arXiv:hep-ph/0406125].
%--------------------------------------------------
\bibitem{Voloshin:2004vk}
  S.~A.~Voloshin,
  %``Parity violation in hot QCD: How to detect it,''
  Phys.\ Rev.\  C {\bf 70}, 057901 (2004).
%  [arXiv:hep-ph/0406311].
%--------------------------------------------------
\bibitem{Kharzeev:2007jp}
  D.~E.~Kharzeev, L.~D.~McLerran and H.~J.~Warringa,
  %``The effects of topological charge change in heavy ion collisions: 'Event by
  %event P and CP violation',''
  Nucl.\ Phys.\  A {\bf 803}, 227 (2008).
%  [arXiv:0711.0950 [hep-ph]].
%--------------------------------------------------
\bibitem{Voloshin:2008jx}
  S.~A.~Voloshin  [STAR Collaboration],
  %``Probe for the strong parity violation effects at RHIC with three particle
  %correlations,''
  arXiv:0806.0029 [nucl-ex].
%--------------------------------------------------
\bibitem{:2009txa}
  B.~I.~Abelev {\it et al.}  [STAR Collaboration],
  %``Observation of charge-dependent azimuthal correlations and possible local
  %strong parity violation in heavy ion collisions,''
  arXiv:0909.1717 [nucl-ex].
%--------------------------------------------------
\bibitem{:2009uh}
  B.~I.~Abelev {\it et al.}  [STAR Collaboration],
  %``Azimuthal Charged-Particle Correlations and Possible Local Strong Parity
  %Violation,''
  Phys.\ Rev.\ Lett.\  {\bf 103}, 251601 (2009).
%  [arXiv:0909.1739 [nucl-ex]].
%--------------------------------------------------
\bibitem{Fukushima:2009ft}
  K.~Fukushima, D.~E.~Kharzeev and H.~J.~Warringa,
  %``Electric-current Susceptibility and the Chiral Magnetic Effect,''
  Nucl.\ Phys.\  A {\bf 836}, 311 (2010).
%  [arXiv:0912.2961 [hep-ph]].
%--------------------------------------------------
\bibitem{Fukushima:2008xe}
  K.~Fukushima, D.~E.~Kharzeev and H.~J.~Warringa,
  %``The Chiral Magnetic Effect,''
  Phys.\ Rev.\  D {\bf 78}, 074033 (2008).
%  [arXiv:0808.3382 [hep-ph]].
%--------------------------------------------------
\bibitem{Warringa:2009rw}
  H.~J.~Warringa,
  %``The Chiral Magnetic Effect: Measuring event-by-event P- and CP-violation
  %with heavy ion-collisions,''
  arXiv:0906.2803 [hep-ph].
%--------------------------------------------------
\bibitem{Asakawa:2010bu}
  M.~Asakawa, A.~Majumder and B.~Muller,
  %``Electric Charge Separation in Strong Transient Magnetic Fields,''
  arXiv:1003.2436 [hep-ph].
%--------------------------------------------------
\bibitem{Buividovich:2009wi}
  P.~V.~Buividovich {\it et al.}, 
  %M.~N.~Chernodub, E.~V.~Luschevskaya and M.~I.~Polikarpov,
  %``Numerical evidence of chiral magnetic effect in lattice gauge theory,''
  arXiv:0907.0494 [hep-lat].
%--------------------------------------------------
\bibitem{Buividovich:2010tn}
  P.~V.~Buividovich, et al.,
  %M.~N.~Chernodub, D.~E.~Kharzeev, T.~Kalaydzhyan,
  %E.~V.~Luschevskaya and M.~I.~Polikarpov,
  %``Magnetic-Field-Induced insulator-conductor transition in SU(2) quenched
  %lattice gauge theory,''
  arXiv:1003.2180 [hep-lat].
%--------------------------------------------------
\bibitem{Nam:2009jb}
  S.~i.~Nam,
  %``Chiral magnetic effect at low temperature,''
  Phys.\ Rev.\  D {\bf 80}, 114025 (2009).
%  [arXiv:0911.0509 [hep-ph]].
%--------------------------------------------------
\bibitem{Nam:2009hq}
  S.~i.~Nam,
  %``Chiral magnetic effect (CME) at low temperature from instanton vacuum,''
  arXiv:0912.1933 [hep-ph], talk given at HNP09, RCNP, Osaka Univ, Japan.
%--------------------------------------------------
\bibitem{Fukushima:2010fe}
  K.~Fukushima, M.~Ruggieri and R.~Gatto,
  %``Chiral magnetic effect in the PNJL model,''
  arXiv:1003.0047 [hep-ph].
%--------------------------------------------------
\bibitem{Fu:2010rs}
  W.~j.~Fu, Y.~x.~Liu and Y.~l.~Wu,
  %``Chiral Magnetic Effect and QCD Phase Transitions with Effective Models,''
  arXiv:1003.4169 [hep-ph].
%--------------------------------------------------
\bibitem{Gorsky:2010xu}
  A.~Gorsky, P.~N.~Kopnin and A.~V.~Zayakin,
  %``On the Chiral Magnetic Effect in Soft-Wall AdS/QCD,''
  arXiv:1003.2293 [hep-ph].
%--------------------------------------------------
\bibitem{Yee:2009vw}
  H.~U.~Yee,
  %``Holographic Chiral Magnetic Conductivity,''
  JHEP {\bf 0911}, 085 (2009).
%  [arXiv:0908.4189 [hep-th]].
%--------------------------------------------------
\bibitem{Rebhan:2009vc}
  A.~Rebhan, A.~Schmitt and S.~A.~Stricker,
  %``Anomalies and the chiral magnetic effect in the Sakai-Sugimoto model,''
  JHEP {\bf 1001}, 026 (2010).
%  [arXiv:0909.4782 [hep-th]].
%--------------------------------------------------
\bibitem{Basar:2010zd}
  G.~Basar, G.~V.~Dunne and D.~E.~Kharzeev,
  %``Chiral Magnetic Spiral,''
  arXiv:1003.3464 [hep-ph].
%--------------------------------------------------
\bibitem{Fukushima:2010zz}
  K.~Fukushima and M.~Ruggieri,
  %``Dielectric correction to the Chiral Magnetic Effect,''
  arXiv:1004.2769 [hep-ph].
%--------------------------------------------------
\bibitem{Schafer:1996wv}
  T.~Schafer and E.~V.~Shuryak,
  %``Instantons in QCD,''
  Rev.\ Mod.\ Phys.\  {\bf 70}, 323 (1998).
%  [arXiv:hep-ph/9610451].
%--------------------------------------------------
\bibitem{Diakonov:2002fq}
  D.~Diakonov,
  %``Instantons at work,''
  Prog.\ Part.\ Nucl.\ Phys.\  {\bf 51}, 173 (2003).
%  [arXiv:hep-ph/0212026].
%--------------------------------------------------
\bibitem{Schafer:1995pz}
  T.~Schafer and E.~V.~Shuryak,
  %``The instanton liquid in QCD at zero and finite $T$,''
  Phys.\ Rev.\  D {\bf 53}, 6522 (1996).
%  [arXiv:hep-ph/9509337].
%--------------------------------------------------
\bibitem{Arnold:1987zg}
  P.~Arnold and L.~D.~McLerran,
  %``The Sphaleron Strikes Back,''
  Phys.\ Rev.\  D {\bf 37}, 1020 (1988).
%--------------------------------------------------
\bibitem{Fukugita:1990gb}
  M.~Fukugita and T.~Yanagida,
  %``SPHALERON INDUCED BARYON NUMBER 
  %NONCONSERVATION AND A CONSTRAINT ON
  %MAJORANA NEUTRINO MASSES,''
  Phys.\ Rev.\  D {\bf 42}, 1285 (1990).
%--------------------------------------------------
\bibitem{Diakonov:1988my}  D.~Diakonov and A.~D.~Mirlin,
  %``INSTANTON VACUUM AT NONZERO $T$S,''
  Phys.\ Lett.\  B {\bf 203}, 299 (1988).
%--------------------------------------------------
\bibitem{Harrington:1976dj}
  B.~J.~Harrington and H.~K.~Shepard,
  %``Euclidean Solutions And Finite $T$ Gauge Theory,''
  Nucl.\ Phys.\  B {\bf 124}, 409 (1977).
%--------------------------------------------------
\bibitem{Diakonov:1995qy}
  D.~Diakonov, M.~V.~Polyakov and C.~Weiss,
  %``Hadronic matrix elements of gluon operators in the instanton vacuum,''
  Nucl.\ Phys.\  B {\bf 461}, 539 (1996).
%  [arXiv:hep-ph/9510232].
%--------------------------------------------------
\bibitem{Schwinger:1951nm}
  J.~S.~Schwinger,
  %``On gauge invariance and vacuum polarization,''
  Phys.\ Rev.\  {\bf 82}, 664 (1951).
%--------------------------------------------------
\bibitem{Nam:2009nn}
  S.~i.~Nam,
  %``An effective thermodynamic potential from the instanton with the 
  % maginary quark-chemical potential,''
  arXiv:0905.3609 [hep-ph], accepted for publication in J.~Phys.~G.
%--------------------------------------------------
\bibitem{Kraan:1998pm}
  T.~C.~Kraan and P.~van Baal,
  %``Periodic instantons with non-trivial holonomy,''
  Nucl.\ Phys.\  B {\bf 533}, 627 (1998).
%  [arXiv:hep-th/9805168].
%--------------------------------------------------
\bibitem{Lee:1998bb}
  K.~M.~Lee and C.~h.~Lu,
  %``SU(2) calorons and magnetic monopoles,''
  Phys.\ Rev.\  D {\bf 58}, 025011 (1998).
%  [arXiv:hep-th/9802108].
%--------------------------------------------------
%--------------------------------------------------
%--------------------------------------------------
%--------------------------------------------------
%--------------------------------------------------
%--------------------------------------------------
%--------------------------------------------------
%--------------------------------------------------
%--------------------------------------------------
%--------------------------------------------------
%--------------------------------------------------
%--------------------------------------------------
%--------------------------------------------------
%--------------------------------------------------
%--------------------------------------------------
%--------------------------------------------------
%--------------------------------------------------
%--------------------------------------------------
%--------------------------------------------------
%--------------------------------------------------
%--------------------------------------------------
%--------------------------------------------------
%--------------------------------------------------
\end{thebibliography}
\end{document}